\documentclass[acmsmall,screen]{acmart}
\AtBeginDocument{%
  }

\setcopyright{acmlicensed}
\copyrightyear{2025}
\acmYear{2025}
\acmDOI{XXXXXXX.XXXXXXX}
\acmJournal{TOSEM}
\usepackage[table]{xcolor} 
\usepackage{tcolorbox}
\usepackage{multirow}
\usepackage{enumitem}
\usepackage{algorithm}
\usepackage{algpseudocode}
\usepackage{graphicx}
\usepackage{booktabs} 
\usepackage[normalem]{ulem}
\usepackage{amsmath}  

\usepackage{amssymb}  
\usepackage{arydshln} 
\newcommand{\G}{\mathbf{G}}  
\newcommand{\F}{\mathbf{F}}  
\usepackage{bbding} 
\usepackage{graphicx} 
\usepackage{booktabs} 
\usepackage{fontawesome}
\usepackage{pifont}
\usepackage{amssymb} 
\usepackage{pifont}  
\usepackage{graphicx} 

\renewcommand{\Until}[1]{\mathbin{\mathbf{U}_{#1}}}

\newcommand{\oomit}[1]{}

{\bfseries}{\rmfamily} 
{\bfseries}{\rmfamily} 
{\bfseries}{\rmfamily}
{\bfseries}{\rmfamily}
\newtheorem{mydefinition}{Definition}{\bfseries}{\rmfamily}
{\bfseries}{\rmfamily}

\begin{document}


\title{ClarifySTL: An Interactive LLM Agent Framework for STL Transformation through Requirements Clarification}
\author{Yue Fang}
\email{y.fang@stu.pku.edu.cn}
\affiliation{%
  \institution{School of Computer Science, Peking University}
  \city{Beijing}
  \country{China}
}

\author{Zhi Jin}
\authornote{Corresponding Author.}
\email{zhijin@pku.edu.cn}
\affiliation{%
  \institution{School of Computer Science, Peking University}
  \city{Beijing}
  \country{China}
}

\author{Jie An}
\authornotemark[1]
\email{anjie@iscas.ac.cn}
\affiliation{%
  \institution{Institute of Software Chinese Academy of Sciences}
  \city{Beijing}
  \country{China}
}

\author{Jia Li}
\email{jia.li@whu.edu.cn}
\affiliation{%
  \institution{School of Computer Science, Wuhan University}
  \city{Wuhan}
  \country{China}
}

\author{Hongshen Chen}
\email{ac@chenhongshen.com}
\affiliation{%
  \institution{JD.com}
  \city{Beijing}
  \country{China}
}

\author{Xiaohong Chen}
\email{xhchen@sei.ecnh.edu.cn}
\affiliation{%
 \institution{East China Normal University}
 \city{Shanghai}
 \country{China}}

\author{Naijun Zhan}
\email{njzhan@pku.edu.cn}
\affiliation{%
  \institution{School of Computer Science, Peking University}
  \city{Beijing}
  \country{China}
}




\renewcommand{\shortauthors}{Yue Fang et al.}

\begin{abstract}
Signal Temporal Logic (STL) is a formal language for specifying real-time behaviors of cyber-physical systems (CPS). Automatically transforming natural language requirements into STL specifications has received growing attention. 
Recent efforts leveraging large language models (LLMs) have demonstrated impressive performance, but some natural language requirements in practice contain vague or ambiguous information, which remains challenging for LLMs to handle.
To address these challenges, we propose ClarifySTL, an interactive LLM-agent framework that enhances STL transformation through requirements clarification. 
ClarifySTL first detects vague expressions that indicate underspecified information in a requirement.
If any vagueness is detected, it generates targeted clarification queries to guide users in supplementing the requirement until all necessary details are provided. 
Subsequently, if ClarifySTL detects ambiguities, it formulates focused ambiguity clarification queries and updates the requirements based on user feedback until all ambiguities are resolved.
Finally, the requirements with vagueness and ambiguity clarified are transformed into STL specifications using LLMs.
This interactive framework ensures that the resulting STL formulas faithfully capture user intent while reducing the burden on the user.
We evaluate ClarifySTL on the representative benchmarks DeepSTL and STL-DivEn, as well as our newly introduced AmbiEval benchmark, which is specifically designed to assess the performance of the agents in handling vagueness and ambiguity, including both detection and query generation. The experimental results show that ClarifySTL outperforms the state-of-the-art model by 13.92\% and 12.08\% in Formula Accuracy and Template Accuracy on the DeepSTL benchmark, and by 12.57\% and 12.99\% in the same metrics on the STL-DivEn benchmark respectively, and achieves an average 90.9\% detection accuracy for vagueness and ambiguity. 
Human evaluations further validate the effectiveness of our framework, which clarifies 93.8\% of defective requirements and generates effective queries for vagueness and ambiguity at rates of 93.3\% and 94.3\%, respectively.

\end{abstract}




\received{20 February 2007}
\received[revised]{12 March 2009}
\received[accepted]{5 June 2009}

\maketitle

\section{Introduction}

Temporal Logic (TL)~\cite{pnueli1977temporal} provides a powerful tool for precisely expressing the dynamic behavior of systems in the analysis and verification of cyber-physical systems (CPS).
Signal Temporal Logic (STL)~\cite{maler2004monitoring}, an extension of TL, introduces real-time and real-valued constraints, enabling the description of not only discrete temporal events but also continuous-time and real-valued dynamic changes.
With these capabilities, STL has been widely used in safety-critical CPS such as autonomous driving and robotic control, helping system designers precisely specify temporal requirements of system behaviors and drawing growing attention from both industry and academia~\cite{maierhofer2020formalization,tellex2020robots,gilpin2020smooth,yan2021stone}.
However, most system requirements with temporal constraints remain documented in informal natural language by domain experts.
Accurately transforming such requirements into formal STL specifications poses a major challenge to the broader application of STL in practical CPS development and verification.

Manually writing precise STL formulas is a laborious task for domain experts, as the process is both time-consuming and error-prone.
Therefore, many studies~\cite{MadsenVSVDWDB18,mao2024nl2stl,chen2023stl} have attempted to develop automated methods to transform natural language descriptions into STL specifications, aiming to reduce the manual burden and improve accuracy.
Existing approaches for transforming natural language requirements into TL and STL specifications can be broadly categorized as rule-based methods and deep learning-based methods. 
Rule-based methods are widely adopted~\cite{LignosRFMK15,ghosh2016arsenal,raman2013sorry,tellex2020robots}, and they use fixed templates to transform natural language into intermediate representations, which are subsequently transformed into temporal logic formulas through manually designed rules.
While these methods reduce human efforts, they not only require extensive expert knowledge and involve a steep learning curve ~\cite{kulkarni2013new}, but also 
are usually limited to highly structured natural language that follows predefined patterns.

With the development of natural language processing, deep learning-based transformation methods have been proposed~\cite{li2023learning,patel2019learning,HeBNIG22,ChenGZF23,Enhancing}.
The representative approaches include the following: DeepSTL~\cite{HeBNIG22} employs a grammar-based data generation technique to train an attentional model that transforms English to STL using transformer-based neural networks. 
Similarly, NL2TL~\cite{ChenGZF23} utilizes LLMs to build the natural language-TL dataset and fine-tunes the T5 model for transformation tasks. 
KGST~\cite{Enhancing} uses a two-stage approach: it first fine-tunes an LLM with synthesized diverse data to transform natural language into an initial STL, then refines it using retrieved examples from an external knowledge base.
These methods provide impressive improvement in the accuracy of transforming natural language into STL specifications, and also enhance the scalability and generalization. 

\begin{figure*}[!t]
    \centering
    \includegraphics[width=1\linewidth]{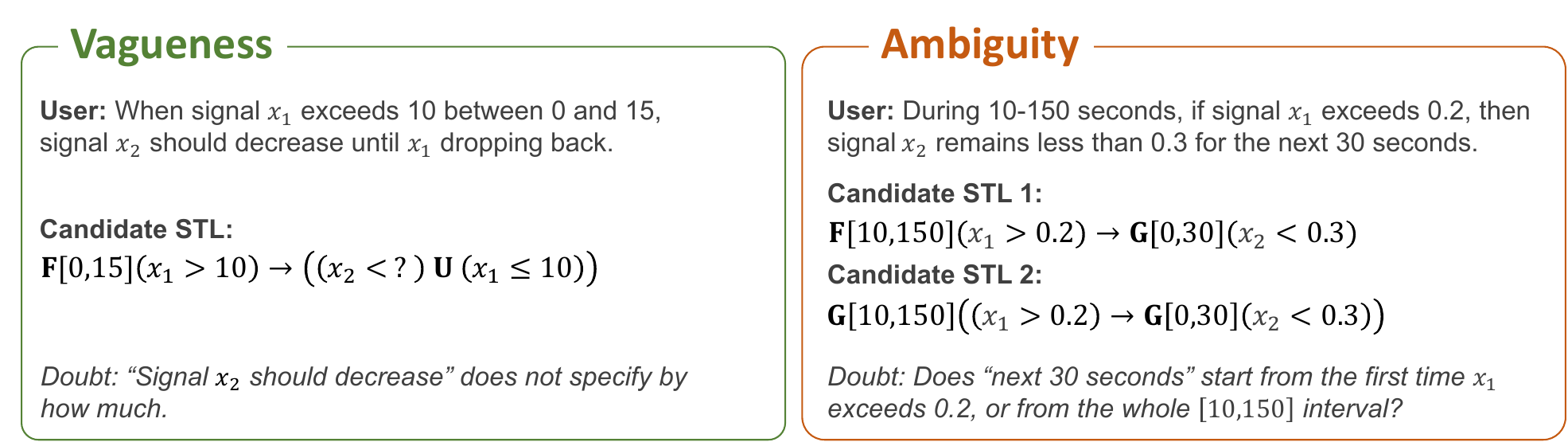}
    \caption{User-written natural language requirements may fail to accurately convey their intent. The left part shows vagueness, where the timing constraint is unspecified.
The right part illustrates ambiguity, where a single requirement can have multiple interpretations.}
    \label{fig:Examples}
\end{figure*}

\begin{figure*}[!t]
    \centering
    \includegraphics[width=1\linewidth]{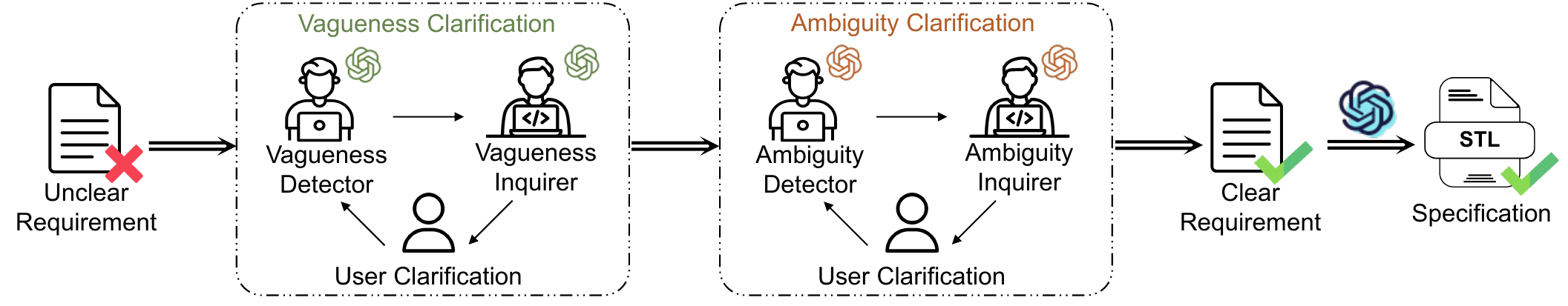}
    \caption{The Brief Structure of ClarifySTL.}
    \label{fig:Brief_structure}
\end{figure*}

However, in practical applications, user-written requirements are often vague or ambiguous due to differences of expertise and perspective among users~\cite{mu2024clarifygpt,2024LLM,thistle1986control}.
Following the requirements engineering literature~\cite{handbook2003contract,berry2004ambiguity,kamsties2000taming,luitel2024improving,ezzini2021using,sabetzadeh21,pragyan2025demystifying}, vagueness refers to requirements with underspecified or imprecisely bounded constraints that need further clarification, whereas ambiguity refers to requirements expressed in a way that admits two or more distinct plausible interpretations.
As shown in Figure~\ref{fig:Examples}, in the case of vagueness, the constraint ``signal $x_1$ should decrease'' lacks specific values. 
In cases of ambiguity, the user input ``within the next 30 seconds'' can be interpreted in two ways: starting from when \( x_1 \) first exceeds 0.2, or within the entire time range from 10 to 150 seconds.
Consequently, directly generating STL specifications from such natural language requirements is hard to satisfy the intent of users.

In this paper, to address these challenges, we propose the ClarifySTL framework, which interactively clarifies vagueness and ambiguity in natural language requirements before their transformation into STL.
The framework identifies vague and ambiguous components in requirements and generates precise clarifying queries, which can not only avoid unnecessary interactions between LLMs and users, but also guide users in providing effective clarifications.
As shown in Figure~\ref{fig:Brief_structure}, the ClarifySTL consists of two stages: vagueness clarification and ambiguity clarification.
In the vagueness clarification stage, given a natural language requirement, we detect its vagueness by training a Vagueness Detector through fine-tuning LLMs on a multi-classification dataset containing requirements with three types of vagueness: temporal, numerical, and conditional logic, 
as well as complete requirements. The detector accurately identifies vague expressions, and the Vagueness Inquirer then generates queries to obtain the underspecified information through Chain-of-Thought~\cite{wei2022chain} (CoT)-based prompting. This process iterates until all essential information is provided.
Even after the information is clarified, the requirements may still face the risk of multiple possible interpretations, requiring the user to determine which one aligns with their intent. 
To address this, during the ambiguity clarification stage, the Ambiguity Detector is trained with contrastive learning on LLMs to identify ambiguous expressions, including referential ambiguity and semantic ambiguity.
Then, the Ambiguity Inquirer generates multiple STL candidates and back-translates them into natural language. It compares these natural language versions with the original requirement, analyzes which parts need clarification, and formulates queries for the user.
This stage is also iterative until the ambiguity is resolved. Finally, the refined requirements are fed into an LLM to generate the STL specifications.

We evaluate our framework on the representative benchmarks DeepSTL and STL-DivEn~\cite{Enhancing}, as well as on AmbiEval, which is proposed in this work and used to assess the detection and query generation capabilities of the agents.
Metric-based evaluation results demonstrate that STL formulas generated from the requirements clarified by our framework achieve higher accuracy than those produced by baselines. 
For example, ClarifySTL outperforms the current state-of-the-art (SOTA) model by 13.92\% and 12.08\% in Formula Accuracy and Template Accuracy on the DeepSTL benchmark, and by 12.57\% and 12.99\% in the same metrics on the STL-DivEn benchmark.
Furthermore, our model obtains an average 90.9\% detection accuracy for vagueness and ambiguity, which surpasses other baselines.
Human evaluation results show that our framework clarifies an average 93.8\% of defective requirements. 
Additionally, 93.3\% of the queries generated by our Inquirer are effective, outperforming other methods.

To systematically evaluate the effectiveness of ClarifySTL, we address the following three research questions:
\noindent\ding{182} \textbf{Transformation Accuracy:} How does ClarifySTL perform compared to existing STL transformation methods in generating accurate STL specifications?
\noindent\ding{183} \textbf{Detection Capability:} How effectively can ClarifySTL detect vagueness and ambiguity in natural language requirements?
\noindent\ding{184} \textbf{Query Quality:} How effective and clear are the clarifying queries generated by ClarifySTL for guiding users to provide accurate clarifications?

In summary, our main contributions are as follows:
\begin{itemize}
\item We propose ClarifySTL, a novel interactive framework that guides users to clarify vagueness and ambiguity in natural language requirements for STL specifications.
\item We introduce vagueness and ambiguity detection modules that accurately identify requirements with the corresponding issues.
\item We propose query generation methods that generate targeted queries for users, guiding them to clarify vagueness and ambiguity in requirements.
\item We conduct extensive experiments on benchmarks, demonstrating that our approach effectively detects issues, generates clarification queries, and enables users to obtain accurate STL specifications.
\end{itemize}

\noindent\textbf{Outline.} In the rest of this paper, Section~\ref{sec:preliminaries} provides the background information. Section~\ref{sec:approach} details our proposed framework, ClarifySTL. Section~\ref{sec:experiments} describes the experimental setup. Section~\ref{sec:results} presents the results and their analysis. Section~\ref{sec:discussion} presents further discussion and shows the threats to validity of ClarifySTL. Section~\ref{sec:related_work} reviews related work in this area. Finally, Section~\ref{sec:conclusion} concludes the paper and outlines directions for future work.

\section{Background: Signal Temporal Logic} \label{sec:preliminaries}
STL is a specification logic widely used  for CPS.
As an illustrative example, consider the \emph{Automatic Transmission (AT)} system, which is frequently employed as a benchmark in system analysis~\cite{ARCHCOMP20Falsification,ARCHCOMP21Falsification,ARCHCOMP22Falsification}.
In automotive engineering, AT serves as a key transmission controller that continuously outputs the vehicle’s \textsf{gear}, \textsf{speed}, and \textsf{rpm}.
One of its safety requirements can be stated as follows:
\emph{In the following 12 seconds, whenever the speed is higher than 45 m/s, the engine speed should remain below 2700 rpm within four seconds.}
STL is capable of expressing such real-time and real-valued constraints.

Let $\mathbb{R}$ stand for the set of real numbers,  $\mathbb{R}_{\geq 0}$ and $\mathbb{R}_+$ respectively for the nonnegative and positive real numbers. 
We use $\mathbb{N}_+$ to denote the set of positive integers. 

Let $T\in \mathbb{R}_+$ be a positive real value, and  $d\in\mathbb{N}_+$ be a positive integer. 
A \emph{$d$-dimensional signal} is defined as a function $\mathbf{v} \colon [0,T] \to \mathbb{R}^d$,
where $T$ represents the \emph{time horizon} of $\mathbf{v}$. 
Given an arbitrary time instant $t\in[0, T]$, $\mathbf{v}(t)$ is a $d$-dimensional real vector; each dimension concerns a signal \emph{variable} that has a certain physical property, e.g., \textsf{speed}, \textsf{rpm}, \textsf{acceleration}, etc. 
In this paper, we select a fixed set $X$ of variables and, without ambiguity, we call a variable a signal ($1$-dimensional signal). 

\begin{mydefinition}[STL Syntax]
\label{def:stl-syntax}
In STL, both atomic formulas $\alpha$ and general formulas $\varphi$ are introduced inductively according to the following rules:
\begin{gather*}
 \alpha \,::\equiv\, f(x_1, \dots, x_K) > 0 \\
 \varphi \,::\equiv\,
\alpha \mid \bot
\mid \neg \varphi 
\mid \varphi_1 \wedge \varphi_2
\mid \G_{I}\varphi
\mid \F_{I}\varphi
\mid \varphi_1 \Until{I} \varphi_2
\end{gather*}
where $f$ is a $K$-ary function $f:\mathbb{R}^K \to \mathbb{R}$, $x_1, \dots, x_K \in X$, and $I$ is a closed non-singular interval in $\mathbb{R}_{\geq 0}$, i.e.,\ $I=[l,u]$, where $l,u \in \mathbb{R}_{\geq 0}$ and $l<u$.
$\G, \F$, and $\Until{}$ are temporal operators, which are known as \emph{always}, \emph{eventually} and \emph{until}, 
 respectively. The always operator $\G$ and eventually the operator $\F$ can be treated as special instances of the until operator $\Until{}$, which can be defined by $\F_{I}\varphi\equiv\top\Until{I}\varphi$ and $\G_{I}\varphi\equiv\lnot\F_{I}\lnot\varphi$.
Other Boolean connectives such as $\lor, \rightarrow$ are provided as syntactic sugar, i.e.,  
$\varphi_1\lor\varphi_2\equiv \neg(\neg\varphi_1\land\neg\varphi_2)$, $\varphi_1\to\varphi_2 \equiv \neg\varphi_1 \lor \varphi_2$. 
\end{mydefinition}

The \emph{Boolean semantics} of an STL formula are defined using a satisfaction relation $(\mathbf{v},t) \models \varphi$, indicating that the signal $\mathbf{v}$  satisfies an STL formula $\varphi$ at the time point $t$:
\begin{align*}
    &(\mathbf{v},t) \models \alpha   &  \Leftrightarrow \quad & f(\mathbf{v}(t))\geq 0 \\
    & (\mathbf{v},t) \models \neg\varphi & \Leftrightarrow  \quad & (\mathbf{v},t) \not\models \varphi \\
   & (\mathbf{v},t) \models \varphi_1 \wedge \varphi_2 & \Leftrightarrow  \quad & (\mathbf{v},t) \models \varphi_1 \wedge (\mathbf{v},t) \models \varphi_2 \\
   & (\mathbf{v},t) \models \G_{[l,u]} \varphi & \Leftrightarrow  \quad & \forall t'\in [t+l,t+u].\, (\mathbf{v},t') \models \varphi \\
   & (\mathbf{v},t) \models \F_{[l,u]} \varphi & \Leftrightarrow  \quad & \exists t'\in [t+l,t+u].\, (\mathbf{v},t') \models \varphi \\
   & (\mathbf{v},t) \models \varphi_1 \Until{[l,u]} \varphi_2 & \Leftrightarrow  \quad & \exists t'\in [t+l,t+u].\, (\mathbf{v},t') \models \varphi_2 
   \wedge \forall t'' \in [t, t']. \, (\mathbf{v},t'') \models \varphi_1
\end{align*}

Now, we are able to formally specify the above \emph{AT safety requirement} using the following STL formula: 
\[\G_{[0,12]}(\textsf{speed} > 45 \to \F_{[1,4]}(\textsf{rpm} < 2700)).\]

It should be noted that a \emph{nested} STL formula is an STL formula in which some temporal operator appears within the scope of other temporal operators.

\section{Approach} \label{sec:approach}
\subsection{Overview}

\begin{figure}[!t]
    \centering
    \includegraphics[width=1\linewidth]{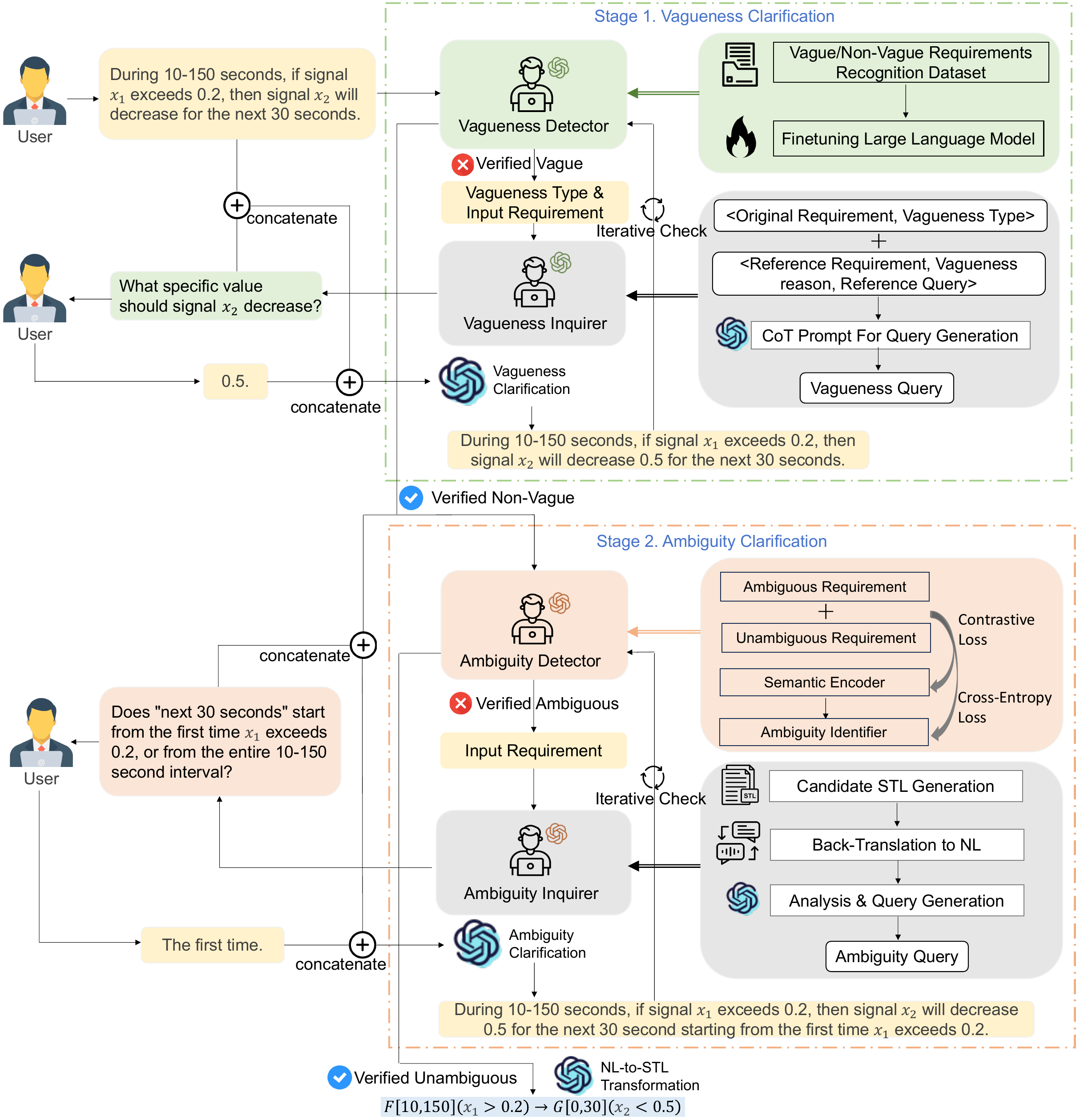}
    \caption{The Overview of ClaritySTL.}
    \label{fig:overviewofClaritySTL}
\end{figure}


The overview of our ClarifySTL model is shown in Figure~\ref{fig:overviewofClaritySTL}. Our approach consists of two stages: vagueness clarification and ambiguity clarification.
Each stage involves a detector and an inquirer. The Vagueness Detector is an LLM fine-tuned on a vagueness detection dataset, and the Vagueness Inquirer generates queries based on CoT prompts; the Ambiguity Detector is built using a contrastive learning-based classifier, and the Ambiguity Inquirer generates clarification queries by analyzing multiple potential interpretations of the input requirement.

After the user inputs a requirement, ClarifySTL first uses the Vagueness Detector to check for vagueness. If vague, it feeds the detected vagueness type and the requirement into the Vagueness Inquirer, which outputs a query to guide the user in clarifying the vagueness. The original requirement, query, and clarification are then sent together to LLMs to generate a supplemented requirement. 
The refined requirement is rechecked by the Vagueness Detector, and the vagueness clarification process repeats until no underspecified information is detected. The process then proceeds to the ambiguity clarification stage.

Subsequently, ClarifySTL uses the Ambiguity Detector to check for ambiguity. If ambiguity is detected, the input requirement is sent into the Ambiguity Inquirer, which generates a query to guide the user in clarifying the ambiguity. The original requirement, query, and user’s clarification are then sent to LLMs to produce a clarified requirement. Similarly, this process is repeated until no ambiguity is detected in the requirement. Finally, ClarifySTL uses an LLM to transform the clarified requirement into an STL specification. 
These iterative re-checking loops serve as a post-clarification verification step that allows the user to confirm whether each refinement preserves the original requirement before the process proceeds to the next iteration.

\noindent\textbf{Example.} As shown in Figure~\ref{fig:overviewofClaritySTL}, when a user inputs the requirement ``During 10--150 seconds, if signal $x_1$ exceeds 0.2, then signal $x_2$ will decrease for the next 30 seconds'' into the vagueness clarification Stage. The Vagueness Detector identifies that this sentence contains underspecified information and categorizes the type as incomplete numerical information. 
It then sends this to the Vagueness Inquirer, which outputs the query ``What specific value should signal $x_2$ decrease?''. 
Subsequently, the user responds with 0.5, which is processed along with the original requirement and the query by the LLM to generate a revised requirement: ``During 10--150 seconds, if signal $x_1$ exceeds 0.2, then signal $x_2$ will decrease 0.5 for the next 30 seconds.''
After a new round of detection confirms the requirement is complete, it proceeds to the next stage. 

After the above vagueness clarification stage, the requirement is first examined by the Ambiguity Detector in the ambiguity clarification stage.
The Ambiguity Detector identifies the ambiguity and then sends the requirement to the Ambiguity Inquirer, which generates a query asking about the start time of ``the next 30 seconds.'' 
After the user clarifies that it means ``the first time'', the clarification, original requirement, and query are sent to the LLM to produce the clarified requirement: ``During 10--150 seconds, if signal $x_1$ exceeds 0.2, then signal $x_2$ will decrease 0.5 for the next 30 seconds starting from the first time $x_1$ exceeds 0.2.'' 
The clarified requirement is verified by the Ambiguity Detector to be unambiguous and is finally transformed by the LLM into the STL formula:  
$\F[10,150](x_1 > 0.2) \rightarrow \G[0,30](x_2 < 0.5)$.

\subsection{Vagueness Detector}

\begin{figure}[!t]
    \centering
    \includegraphics[width=1\linewidth]{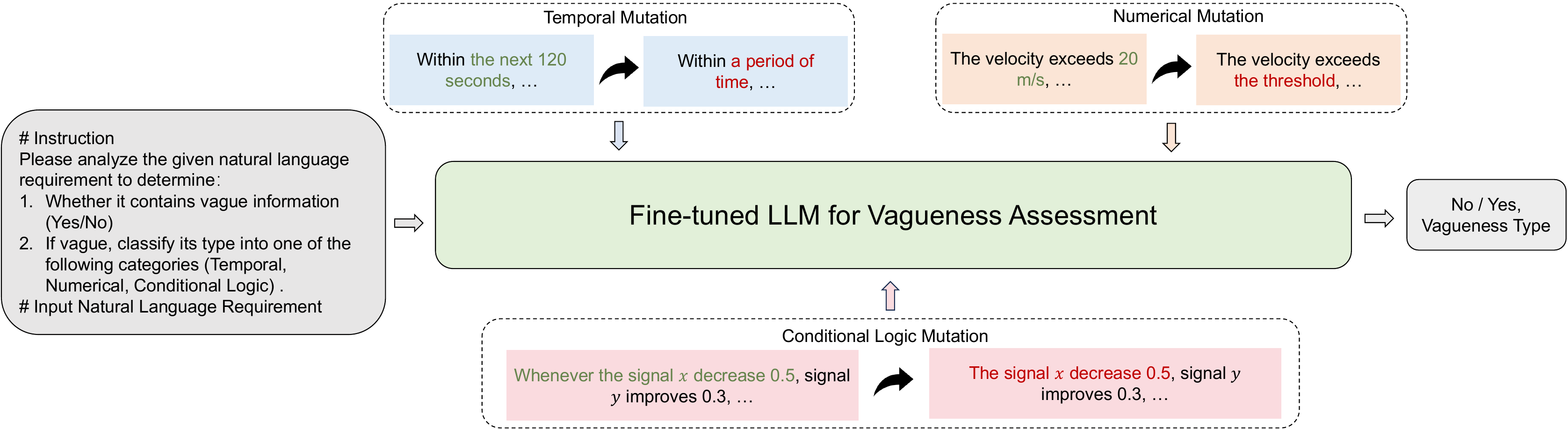}
    \caption{Fine-tuning the Vagueness Detector for Identification.}
    \label{fig:Method1}
\end{figure}

As LLMs lack domain knowledge about vagueness in STL-relevant expressions, they often misjudge precise requirements as vague while overlooking genuinely vague descriptions.
To address this issue, we categorize the types of vagueness and fine-tune LLMs on datasets constructed according to these categories, enabling LLMs to accurately identify vague information.
Since STL syntax in Definition~\ref{def:stl-syntax} is built from temporal intervals, numerical predicates, and logical compositions, we define vagueness by the underspecified information needed to instantiate these STL components. Accordingly, for NL-to-STL transformation, we categorize vagueness into the following three types:

\begin{enumerate}[leftmargin=*]
    \item \textbf{Temporal Information Vagueness.} Temporal constraints in natural language requirements, such as duration, start time, or time intervals, are often partially omitted or not expressed with explicit temporal information, but instead are conveyed using terms like ``soon'', ``later'', or ``in a moment''~\cite{mohammadinejad2024systematic}. This lack of precision makes it impossible to accurately construct precise temporal constraints in the corresponding STL formulas.
    \item \textbf{Numerical Information Vagueness.} Expressions involving numerical values (e.g., rpm, speed, temperature) are often omitted or replaced by qualitative descriptors in natural language requirements~\cite{mohammadinejad2024systematic}, resulting in missing numeric thresholds in the corresponding STL formulas and making it impossible to fully define the quantitative aspects of system behavior.
    \item \textbf{Conditional Logic Information Vagueness.} Conditional logic in natural language is often partially omitted or insufficiently expressed. This vagueness reduces the precision of conditional judgments and makes it difficult to infer the correct logical operators during transformation into STL formulas, affecting the accuracy and effectiveness of the resulting formal specification.
\end{enumerate}

Subsequently, we construct the AmbiEval dataset, which consists of two parts. Here, we introduce Part A of the dataset, which contains vague requirements and their types of incompleteness, and the corresponding complete requirements. 
For each selected input from DeepSTL and STL-DivEn, we first ask the teacher model to identify STL-relevant elements that can be perturbed according to our vagueness taxonomy. If multiple valid perturbation opportunities exist in the same requirement, the teacher model processes them iteratively and introduce multiple vague elements into the same mutated requirement. Each mutated requirement therefore contain more than one type of vague information. All valid mutated requirements are retained in AmbiEval.

Specifically, we employ GPT-4o as the teacher model and guide it with carefully designed mutation prompts to rewrite the original requirements in a directed, type-specific manner. 
To ensure linguistic validity, we require that the teacher model ``keep grammatical correctness and semantic coherence'' in prompts, and we further implement a rule-based syntactic validator that detects grammatical issues and automatically filters out mutants that fail the check. Let $V_{\mathrm{T}}$, $V_{\mathrm{N}}$, and $V_{\mathrm{C}}$ denote the expression sets used for temporal, numerical, and conditional-logic vagueness, respectively. For each mutation, $w$ denotes a vague or underspecified expression sampled from the corresponding set.

\begin{itemize}[leftmargin=*]
\item \textbf{Temporal Mutation.} Remove or replace the time interval in the requirement with a vague description. Formally, when the corresponding STL formula contains a bounded temporal operator $G_{[a,b]}$, $F_{[a,b]}$, or $U_{[a,b]}$, the interval $[a,b]$ in the aligned span of the requirement is either deleted or replaced by a vague phrase $w \in V_{\mathrm{T}}$ (e.g., replace an explicit ``[0, 10] time units'' with ``within the next period of time'').

\item \textbf{Numerical Mutation.} Replace numeric thresholds or concrete values in the requirement with vague descriptions, or delete them directly. Formally, when the corresponding STL formula contains an atomic predicate $f(x_1,\dots,x_K) \sim c$ with $c \in \mathbb{R}$, the threshold $c$ in the aligned span is either deleted or replaced by a qualitative descriptor $w \in V_{\mathrm{N}}$ (e.g., change ``speed exceeds 50'' to ``speed is high'').

\item \textbf{Conditional Logic Mutation.} Delete the logical or conditional relationship in the requirement or replace it with an underspecified expression. Formally, when the corresponding STL formula contains a logical connective $\circ \in \{\neg, \wedge, \vee, \rightarrow\}$, the expression that corresponds to $\circ$ in the requirement is either deleted or replaced by an underspecified expression $w \in V_{\mathrm{C}}$ (e.g., change ``if speed $> 50$ then brake activates'' to ``speed $> 50$, brake activates'').
\end{itemize}


We then fine-tune an LLM on this dataset to obtain an Vagueness Detector capable of identifying which requirements contain underspecified information, as shown in Figure~\ref{fig:Method1}.
Specifically, each training pair in the dataset is structured with three attributes: \texttt{Instruction}, \texttt{Input}, and \texttt{Output}.
The \texttt{instruction} provides the guidance to determine whether a natural language requirement is incomplete and to classify the type of vagueness.
The \texttt{input} is the natural language requirement to be evaluated and is provided to the LLM together with the instruction.
The LLM is trained to generate the judgment result of the input as the \texttt{output}. After training, the model functions as the Vagueness Detector.


\subsection{Vagueness Inquirer}

\begin{figure}[!t]
    \centering
    \includegraphics[width=1\linewidth]{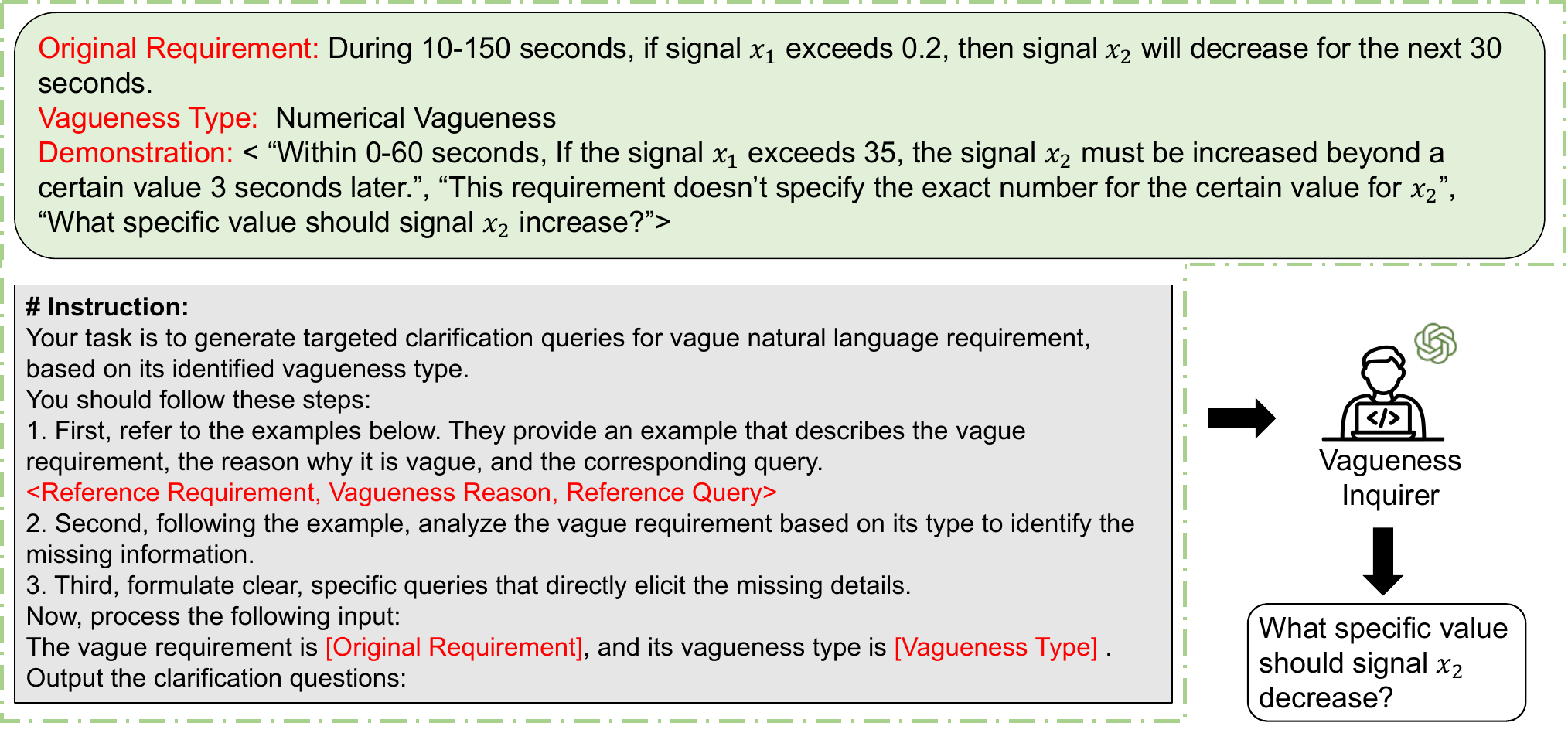}
    \caption{Prompts for the Vagueness Inquirer to Generate Clarification Queries.}
    \label{fig:Incompleteness_Inquerier}
\end{figure}

Precise clarification questions help users clearly express their intentions and ensure the information targets the incomplete parts of the requirements.
Vague or overly broad questions may confuse users, produce irrelevant answers, and increase their burden, making it difficult to provide effective clarifications or causing them to give incorrect ones. ~\cite{mu2024clarifygpt}.
Therefore, once vague requirements are identified, it is crucial to equip LLMs with the ability to generate precise questions. 

To achieve this, we design a Vagueness Inquerier to 
automatically generate targeted clarification questions for vagueness clarification. 
The agent leverages a CoT prompting strategy, enabling step-by-step reasoning over vague requirements. 
As shown in Figure~\ref{fig:Incompleteness_Inquerier}, the prompt consists of three components: Instruction, Requirement Input, and Demonstrations.
\begin{enumerate}[leftmargin=*]
\item \textbf{Instruction.} Specifies the task for the model, i.e., generating clarification questions based on vague requirements and their error types.
\item \textbf{Requirement Input.} Provides the vague user requirements along with their vagueness types to the model.
\item \textbf{Demonstrations.} Selects examples corresponding to the requirement’s error type, enabling the model to learn in context from <Reference Requirement, Incompleteness Reason, Reference Query> triples and to formulate targeted questions that elicit the underspecified information needed to construct complete STL formulas.
    
    

\end{enumerate}

By structuring the agent in this way, it can effectively leverage the reasoning and text generation capabilities of LLMs to address multiple forms of natural language vagueness, ensuring that the generated questions are both specific and informative.

To prevent the Vagueness Inquirer from altering the user’s intent, we explicitly add a constraint in its prompt that requires the generated clarification questions to focus only on the vagueness type and the missing components reported by the Vagueness Detector, and not to introduce information unrelated to the detected vague segments of the requirement. In addition, the prompt allows the user to indicate that the requirement does not contain vagueness if a non-vague sentence is mistakenly fed to the Vagueness Inquirer. In such cases, the agent directly returns the original requirement without adding any supplementary information and passes it to the next stage. When the user's answer fails to effectively supplement the targeted information, the system asks the user to restate the intent.

\subsection{Ambiguity Detector}






Ambiguous and unambiguous natural language may exhibit high similarity in syntactic structure and lexical expressions, such as similar sentence patterns, frequently used terms, or descriptive styles, which makes distinguishing them challenging. 
The main difficulty lies in capturing their representative features from subtle differences in word usage and semantics. 
Therefore, we propose to encode both ambiguous and unambiguous requirements using contrastive learning and subsequently classify them based on their encoded representations. 

Hoffer and Ailon~\cite{2015Deep} introduce the triplet network as a method for contrastive representation learning. 
The network input is a triplet $(x, x^+, x^-)$, where $x$ is a given natural language sentence (i.e., the anchor), $x^+$ is a positive sample from the same class, and $x^-$ is a negative sample from a different class. 
The triplet network learns useful representations for downstream classification tasks by minimizing the distance between $x$ and $x^+$ while maximizing the distance between $x$ and $x^-$. 
In order to more effectively encode representations of ambiguous and unambiguous requirements, we adopt a triplet network-based contrastive learning approach to build the Ambiguity Detector. 

Ambiguity in NL-to-STL transformation mainly affects either the referenced entity or the interpretation of temporal, logical, and relational constraints; accordingly, we categorize ambiguity into two types: referential ambiguity and semantic ambiguity.
We use unambiguous requirements from the DeepSTL and STL-DivEn datasets and, following the same mutation pipeline as in Section~3.2, introduce controlled ambiguities by prompting GPT-4o under two type-specific mutation operators to generate their ambiguous variants:

\begin{itemize}[leftmargin=*]
\item \textbf{Referential Ambiguity Mutation.} Replace an explicit signal name in the requirement with a pronoun or a generic noun phrase, so that the signal referent becomes underdetermined. 
Formally, when the corresponding STL formula contains $K \ge 2$ distinct signal identifiers $\{x_1, \dots, x_K\}$, an occurrence of a signal identifier in the aligned span is replaced by a referring expression $w \in V_{\mathrm{R}}$, where $V_{\mathrm{R}}$ denotes the set of pronouns and generic noun phrases used for referential ambiguity and $w$ is a sampled expression from this set (e.g., change ``$x_1$ exceeds 0.2 and $x_1$ stays high'' to ``$x_1$ exceeds 0.2 and it stays high'').

\item \textbf{Semantic Ambiguity Mutation.} Rewrite the requirement so that the involved signals and predicates remain unchanged, but the sentence admits multiple plausible STL interpretations.
This mutation is realized through in-context learning: we prompt GPT-4o with a fixed set of demonstration triples \emph{(original requirement, ambiguous rewrite, ambiguity explanation)} and ask it to produce a similar rewrite for each new input requirement. 
For example, ``$x_2$ should stay below 0.5 within 30 seconds after $x_1$ exceeds 0.2'' can be rewritten as ``within the next 30 seconds, if $x_1$ exceeds 0.2, then $x_2$ should stay below 0.5'', where ``the next 30 seconds'' admits both ``starting from when $x_1$ exceeds 0.2'' and ``starting from the current moment''. The full prompts are provided in the supplementary file.
\end{itemize}
The generated mutant is first filtered by syntactic validator and then manually validated by two annotators with at least three years of experience in writing STL specifications, who independently verify that the mutant indeed admits multiple plausible STL interpretations.
This forms Part B of the AmbiEval, which includes the ambiguous natural language requirements, their corresponding unambiguous requirements, and labels, where ambiguous requirements are labeled as 1 and unambiguous ones as 0.


\begin{figure}[!t]
    \centering
    \includegraphics[width=0.85\linewidth]{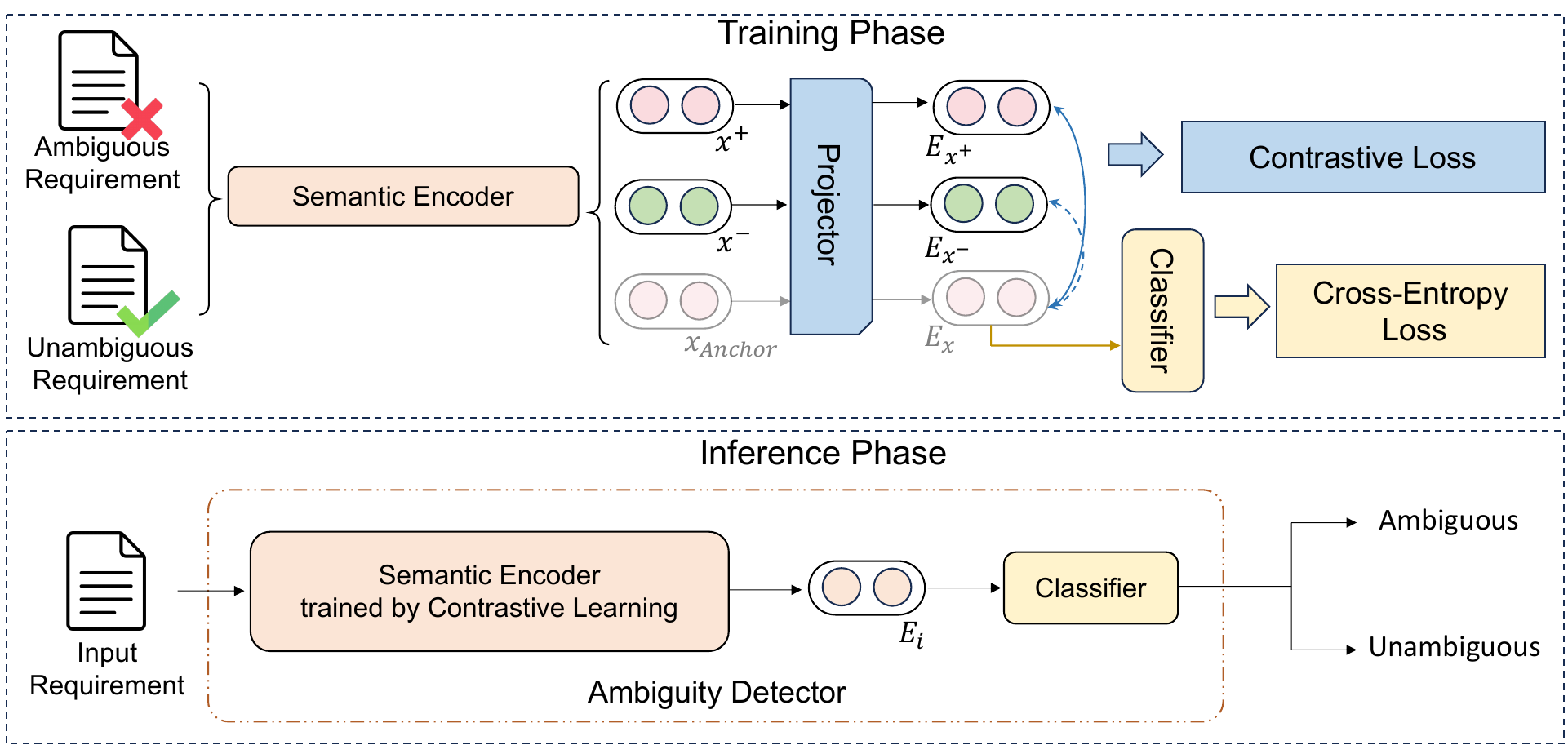}
    \caption{
    Training and Inference Phases of the Ambiguity Detector.
    }
    \label{fig:Ambiguity_Detector}
\end{figure}

As shown in Figure~\ref{fig:Ambiguity_Detector}, for a given sentence $x$, we assume that $x^+$ belongs to the same class as $x$, while $x^-$ belongs to a different class. 
Specifically, if $x$ is an ambiguous requirement, $x^+$ is also ambiguous, whereas $x^-$ is unambiguous. 
Our model uses the hidden states of LLaMA 3-8B as the base model to obtain embeddings for $x$, $x^+$, and $x^-$. 
To perform contrastive learning, we first follow prior works~\cite{2021SimCSE,2024Distinguishing} and forward the same input twice with different hidden-layer dropout masks (dropout probability $=0.1$) to generate $x$ and $x^+$. 

Specifically, the original sentence $x$ is encoded twice with dropout using LLaMA 3-8B as the semantic encoder. We extract the hidden states from its last transformer layer and apply mean pooling over all token representations to obtain two distinct sentence-level representations, which serve as the anchor $x$ and the positive instance $x^+$ for contrastive learning~\cite{2015Deep}. During this process, the parameters of the semantic encoder are frozen in order to preserve the pretrained semantic knowledge, and only the projector and the subsequent lightweight classification head are updated. Subsequently, we use a structure called a \textit{projector} to transform the outputs of the semantic encoder into final embeddings $E_x$, $E_{x^+}$, and $E_{x^-}$. A contrastive loss is then computed based on the cosine distance of the embeddings to train the network. The training objective is to minimize the distance between $x$ and $x^+$ while separating $x$ and $x^-$, which is formulated as:
\[
\mathcal{L}_{\text{c}} = \max \big( \|E_x - E_{x^+}\| - \|E_x - E_{x^-}\| + \epsilon, 0 \big)
\]
where $\epsilon$ denotes the margin between $x$ and $x^-$, with a default value of 1.



By using contrastive learning, the Ambiguity Detector is able to learn meaningful representations $E_x$ for an input natural language requirement $x$. These embeddings are then passed through a classification layer with the softmax function to obtain the probability for each class. A threshold of 0.5 is applied to decide whether the sentence is ambiguous or unambiguous.

\subsection{Ambiguity Inquirer}
\begin{figure}
    \centering
    \includegraphics[width=1\linewidth]{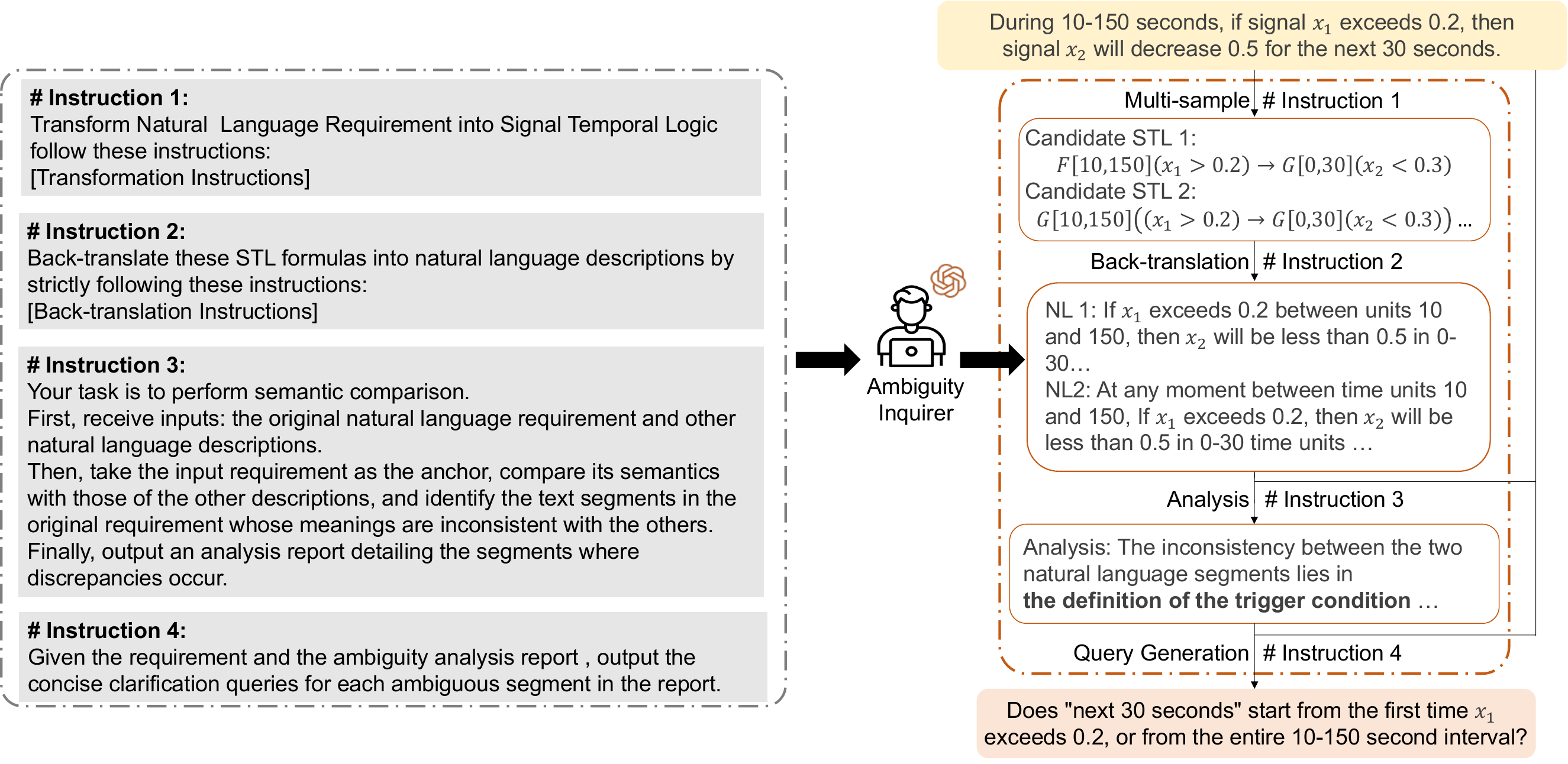}
    \caption{Workflow and Prompts for Ambiguity Inquirer to Generate queries about Ambiguity Clarification.}
    \label{fig:ambiguity_inquirer}
\end{figure}

When the natural language requirements are identified as ambiguous, we generate targeted queries to guide the user to provide clarifications.
To avoid asking the user to rewrite the entire requirement, we design a workflow that leverages LLMs to locate ambiguous parts and generate corresponding clarification queries, as shown in Fig.~\ref{fig:ambiguity_inquirer}. 
First, we leverage LLMs to perform multiple samplings of the input original requirement to obtain STL specification candidates. 
These candidates are back-translated into natural language, and LLMs analyze the differences between them and the original requirement to generate a report. 
Finally, based on this report, the LLM formulates clarification queries. 
The prompts for STL transformation and back-translation can be found in the supplementary file.
The details are as follows:

\begin{itemize}[leftmargin=*]
\item \textbf{Candidate STL Generation.} To better help the LLMs locate ambiguous portions, we first sample the same original requirement multiple times under the same prompt and parameters, generating a set of candidate STL formulas \(\{\phi_1, \phi_2, \dots, \phi_N\}\), representing different possible interpretations of the same requirement. 
By comparing the differences among these specifications, potential points of ambiguity in the requirement can be preliminarily located. 
\item \textbf{Back-Translation to Natural Language.} 
Since LLMs have limited capability for direct symbolic-level analysis~\cite{chencodesteer}, we back-translate the STL candidates into the corresponding natural language descriptions \(\{d_1, d_2, \dots, d_N\}\) and leverage the strong reasoning capabilities of LLMs in natural language to conduct analysis. 
To minimize interference from lexical variability, each prompt enforces a fixed translation scheme for variables and symbols, ensuring that the generated descriptions are uniform and facilitating semantic comparison by LLMs.
\item \textbf{GPT-Based Difference Analysis.} 
The original requirement \(R\) and all candidate descriptions \(\{d_1, d_2, \dots, d_N\}\) are fed into an LLM, which performs a comprehensive semantic comparison across all input natural language descriptions and generates a discrepancy analysis report. 
The report will reveal the differences within these natural language descriptions, enabling the model to locate which parts of the original requirement \(R\) may introduce ambiguity.
\item \textbf{Clarification Query Generation.} 
Finally, we provide the original natural language requirement and the discrepancy analysis report to the LLMs, prompting them to generate precise queries for the user to clarify the ambiguity in the requirement.
\end{itemize}

We further add a constraint in the prompt that restricts the generated queries to the divergence points identified in the discrepancy analysis report, so that the queries do not introduce information unrelated to the detected ambiguous segments. In addition, the prompt allows the user to indicate that the requirement does not contain ambiguity if an unambiguous sentence is mistakenly provided to the Ambiguity Inquirer. In such cases, the agent directly returns the original requirement without modifying any content and passes it to the next stage. When the user's answer fails to effectively clarify the ambiguous part, the system asks the user to restate the intent.
\section{Study Design} \label{sec:experiments}
To evaluate the effectiveness of our ClarifySTL\footnote{Supplementary file and project are available at: \url{https://zenodo.org/records/17561877}}, we design three research questions (RQ) and
perform studies to answer these questions. In this section, we describe the research
questions and details of our study. We present the evaluation results and analysis in Section~\ref{sec:results}. 

\subsection{Research Questions}
We address the following three research questions to systematically evaluate the performance and effectiveness of ClarifySTL.

\textbf{RQ1: How does ClarifySTL perform compared to existing STL transformation methods?}
This research question aims to evaluate whether ClarifySTL generates more accurate and semantically consistent STL specifications than existing methods.
To achieve this, we compare ClarifySTL against two types of methods: 11 baselines that directly transform natural language into STL, and five LLM–based interactive transformation methods.
The comparison is conducted on two representative STL transformation benchmarks, including DeepSTL and STL-DivEn.

\textbf{RQ2: How effective is ClarifySTL at detecting vagueness and ambiguity in natural language requirements?}
The ability to detect vagueness and ambiguity is critical for ensuring the reliability of ClarifySTL framework.
In this research question, we investigate ClarifySTL’s capability to identify vague and ambiguous requirements.

\textbf{RQ3: Are the queries generated by ClarifySTL effective and clear for users to provide clarification?}
Generating high-quality clarification queries can significantly reduce the user's effort.
In this research question, 
we evaluate whether the queries generated by our framework accurately highlight the vague or ambiguous portions of the requirement and whether they are easy for users to understand.

\subsection{Datasets}
Our datasets include two representative Natural Language-to-Signal Temporal Logic (NL-to-STL) datasets and one dataset designed for detection and query generation tasks.
Specifically, we adopt two NL-STL datasets, DeepSTL~\cite{HeBNIG22} and STL-DivEn~\cite{Enhancing}, to evaluate the model’s ability to perform NL-to-STL transformation.
Additionally, in this work, we propose the AmbiEval dataset. It is used to fine-tune models for detecting vagueness and ambiguity, as well as to evaluate ClarifySTL’s performance in identification and query generation.

\begin{itemize}[leftmargin=*]
\item \textbf{DeepSTL} is a synthetic dataset generated by rule-based methods. To ensure systematic coverage, DeepSTL defines a structured fragment of STL formulas with three layers and performs random sampling according to empirical operator distributions. Each sampled STL formula is then translated into multiple natural language requirements using a template-based translation procedure, enriched with synonymous expressions to capture linguistic variability.
\item \textbf{STL-DivEn} is a dataset designed to enhance the diversity and quality of NL-STL pairs. STL-DivEn begins with a handcrafted seed set of 120 NL-STL pairs covering both basic and nested logic. Representative samples are selected through clustering and used to guide LLMs in generating new pairs. The generated data are refined using rule-based filters and human validation, and the validated pairs are iteratively added to expand the dataset.  
\item \textbf{AmbiEval} is a dataset adapted from DeepSTL and STL-DivEn, used to fine-tune LLMs for detecting vagueness and ambiguity and to evaluate the performance of models in detection and query generation. It contains 9,200 labeled instances: Part A includes 6,000 instances for vagueness detection across three types of vagueness, and Part B includes 3,200 instances for ambiguity detection, covering 1,600 instances of referential ambiguity and 1,600 instances of semantic ambiguity. Each vague instance has a corresponding non-vague version, and each ambiguous instance has a corresponding unambiguous version. Each instance is also paired with a human-written query, serving as the ground truth for query generation.

\end{itemize}
\begin{table}[]
\small
\caption{Number of Requirements with Vagueness and Ambiguity Identified in the DeepSTL and STL-DivEn Datasets. ``Both'' Denotes Requirements Containing Both Vagueness and Ambiguity.}
\resizebox{0.9\textwidth}{!}{
\begin{tabular}{lcccccc}
\hline
\multirow{2}{*}{Dataset} & \multicolumn{3}{c}{Vagueness} & \multicolumn{2}{c}{Ambiguity} & \multirow{2}{*}{Both} \\ \cmidrule(lr){2-4} \cmidrule(lr){5-6}
 & Temporal & Numerical & Conditional Logic & Referential & Semantic &  \\ \hline
DeepSTL & 87 & 39 & 103 & 83 & 117 & 62 \\
STL-DivEn & 67 & 56 & 43 & 71 & 212 & 77 \\ \hline
\end{tabular}
}
\label{Table_1}
\end{table}

We randomly select 14,000 samples from DeepSTL and STL-DivEn as the training set and 2,000 samples as the test set for NL-to-STL evaluation. To support the human evaluation in RQ1, we further identify requirements containing vagueness or ambiguity from the NL-to-STL evaluation data. Table~\ref{Table_1} summarizes the distribution of these selected defective requirements. In DeepSTL, the selected subset contains 87 temporal, 39 numerical, and 103 conditional-logic vagueness instances, as well as 83 referential and 117 semantic ambiguity instances, with 62 requirements containing both vagueness and ambiguity. In STL-DivEn, the selected subset contains 67 temporal, 56 numerical, and 43 conditional-logic vagueness instances, together with 71 referential and 212 semantic ambiguity instances, with 77 requirements containing both types of issues.

AmbiEval is constructed separately for training and evaluating the vagueness and ambiguity detectors, as well as the query generation modules. To prevent data leakage, AmbiEval is constructed exclusively from the training split of DeepSTL and STL-DivEn and has no overlap with the 2,000 test samples used for NL-to-STL evaluation. For AmbiEval, we perform a random selection with an even distribution across each category in Part A, resulting in 5,400 samples for training and 600 for testing. Similarly, in Part B, 2,800 samples are selected for training and 400 for testing.



\subsection{Evaluation Metrics}
Following previous works~\cite{HeBNIG22,Enhancing}, we adopt formula accuracy, template accuracy, and BLEU~\cite{papineni2002bleu} as evaluation metrics to assess the performance of NL-to-STL.
Before computing these metrics, all generated STL formulas are first checked by a standard STL syntax checker, and formulas that fail to pass the STL syntax checker are directly considered incorrect.

Formula Accuracy measures token-level alignment between the generated STL formula and the reference formula. Specifically, we split both formulas into STL tokens, compare the tokens at the same positions, and compute the proportion of reference tokens correctly matched by the generated formula.
Template Accuracy first extracts both formulas into STL templates by replacing concrete signal names, predicates, numerical thresholds, and time bounds with template symbols, and then computes the same token-level alignment score over the resulting templates.
BLEU is originally designed for machine translation and evaluates n-gram overlap, reflecting lexical semantic similarity.



In addition to string-level metrics, STL semantic robustness provides a complementary way to check the semantic consistency of the generated formulas. 
In our setting, for each ground-truth STL formula, we first generate a set of signal traces according to its variables, temporal horizons, and numerical thresholds. These traces are sampled to include satisfying and violating cases with respect to the ground-truth formula. The semantic robustness score is then computed as the percentage of traces on which the generated formula and the ground-truth formula are either both satisfied or both violated.

For vagueness and ambiguity detection tasks, i.e., classification tasks, we use Accuracy, Precision, Recall, and F1-score. Accuracy measures the proportion of correctly predicted instances. Precision measures the proportion of correctly predicted positive instances among all instances predicted as positive, while Recall measures the proportion of actual positive instances that are correctly identified. F1-score is the harmonic mean of precision and recall, providing a balanced measure of model performance. 

For query generation tasks, we use ROUGE~\cite{lin2004rouge} and BERTScore~\cite{zhangbertscore}. ROUGE evaluates the overlap between generated text and reference text, commonly used in text generation tasks. BERTScore computes the semantic similarity between generated text and reference text using contextual embeddings from pre-trained language models.

\subsection{Baselines} \label{sec:baselines}
We compare ClarifySTL with 16 baseline methods, which are grouped into direct transformation and interactive transformation.

\subsubsection*{Direct Transformation (Open-source Models)}
\begin{itemize}[leftmargin=*]
    \item \textbf{DeepSTL} is the pioneering work that first applied deep learning to learn STL transformations from NL-to-STL datasets.

    \item \textbf{KGST} is the current SOTA model, featuring a two-stage framework: it fine-tunes an LLM on NL-to-STL data to generate preliminary STL formulas, then refines them using retrieved reference examples through LLMs.

    \item \textbf{Qwen 3-8B} and \textbf{Qwen 3-14B}~\cite{yang2025qwen3} are LLMs based on a Mixture-of-Experts architecture with optimized inference efficiency, demonstrating strong instruction-following capabilities.

    \item \textbf{LLaMA 3-8B} and \textbf{LLaMA 3-14B}~\cite{dubey2024llama} are general-purpose models trained on large-scale data, excelling in text generation and reasoning.
\end{itemize}

\subsubsection*{Direct Transformation (Closed-source Models)}
\begin{itemize}[leftmargin=*]
    \item \textbf{DeepSeek-V3}~\cite{liu2024deepseek} is an advanced Mixture-of-Experts model employing multi-stage training to achieve strong reasoning and text generation performance.

    \item \textbf{GPT-4o}~\cite{hurst2024gpt} is one of the most powerful LLMs, achieving SOTA results in formal reasoning tasks.

    \item \textbf{GPT-4o-mini} is a compact version of GPT-4o, optimized for speed and efficiency while retaining most reasoning capabilities.

    \item \textbf{Gemini-2.5-Pro}~\cite{comanici2025gemini} is a large language model excelling in reasoning and text generation tasks.

    \item \textbf{Claude-4-Sonnet}~\footnote{https://www.anthropic.com/news/claude-4} is a high-performance version of the Claude-4 series, performing reliably on formal reasoning tasks.
\end{itemize}

Among the direct transformation baselines, DeepSTL and KGST follow their original implementations.
For the remaining LLM-based baselines (both open-source and closed-source), we adopt few-shot prompts that include representative NL-STL pairs as in-context examples, along with explicit task instructions and the STL grammar definition, to guide the models in performing the transformation.
To ensure fair comparison, we set the temperature to 0 for deterministic outputs. All prompt templates are provided in the supplementary file.

\subsubsection*{Interactive Transformation}
Interactive transformation follows the same two-stage workflow (vagueness clarification followed by ambiguity clarification) as ClarifySTL, but replaces ClarifySTL's specialized modules with general-purpose LLM capabilities.
Specifically, ClarifySTL employs a fine-tuned Vagueness Detector, a contrastive learning-based Ambiguity Detector, and task-specific query generation strategies (CoT-based prompting for vagueness and back-translation-based analysis for ambiguity).
In contrast, the interactive transformation baselines use a single unified few-shot prompt to instruct LLMs to perform all tasks, including detecting vagueness and ambiguity and generating clarification queries, relying solely on the intrinsic knowledge of LLMs without any fine-tuned detectors or specialized strategies.
In our preliminary experiments, we find that open-source models yield poor performance under this setting; therefore, we only use the aforementioned five closed-source models as interactive transformation baselines.

\subsection{Implementation Details} \label{sec:implementation_details}
Our experiments are conducted on 8 NVIDIA A100 GPUs with 40GB of memory. We implement our framework using PyTorch~\cite{paszke2019pytorch}, Huggingface Transformers~\cite{wolf2020transformers}, and LLaMA-Factory~\cite{zheng2024llamafactory}. 
In ClarifySTL, GPT-4o is used for vagueness query generation, ambiguity query generation, and requirement refinement.
Our framework uses four LLaMA 3-8B models. 
The Vagueness Detector and the two NL-to-STL transformation models are fine-tuned on the corresponding datasets, while the Ambiguity Detector uses a frozen LLaMA 3-8B encoder together with a trainable projector and a lightweight classification head. The trainable models are optimized for 10 epochs using Adam~\cite{kingma2014adam}, with a learning rate of 5e-5 and a batch size of 16.

For the Ambiguity Detector, the projector is implemented as a two-layer MLP with ReLU activation, whose dimensions are 4096$\rightarrow$1024$\rightarrow$256, followed by L2 normalization.
For detection tasks, we use few-shot prompts that include the task definition (i.e., identifying whether a requirement is vague or ambiguous), the category descriptions, and representative labeled examples for each category. We experiment with varying numbers of in-context examples and find that 3 examples yield the best performance across all models, which we adopt as the default configuration. 
For clarifying query generation, the maximum number of generated tokens is set to 300. The temperature is set to 0 by default, except when sampling candidate STL formulas in ambiguity clarification, where it is set to 0.9.

For KGST, we fine-tune LLaMA 3-8B separately on DeepSTL and STL-DivEn, and use GPT-4o to refine the preliminary STL formulas. The training hyperparameters are the same as those used in ClarifySTL. For the interactive transformation baselines, we use a unified few-shot prompt to guide LLMs in defect detection and query generation. The prompts for transformation, requirement refinement, and interactive transformation are provided in the supplementary file.
\section{Results and Analyses} \label{sec:results}
In this section, we answer the three research questions based on our experimental results.

\subsection{RQ1: How does ClarifySTL perform compared to existing STL transformation methods?}

To evaluate the effectiveness of ClarifySTL, we design experiments to compare its performance with 16 baselines in the NL-to-STL task, using DeepSTL and STL-DivEn as benchmarks. The experimental setup is detailed in Sections 4.4 and 4.5.

To simulate a realistic feedback process, we recruit 10 participants, all of whom are graduate students or researchers in the field of CPS with at least three years of experience in writing and interpreting STL specifications.
Before the evaluation, each participant receives a guideline document that describes the task objective, the format of the clarification queries, and examples of expected responses.
Each participant is provided with a set of requirements and automatically generated clarification queries, and is asked to answer these queries.
Each defective requirement is clarified by three participants, and the experimental results are reported as the average of all responses.
To assess the consistency among participants, we evaluate agreement from two perspectives.
First, for the correctness judgments of the final STL formulas, we compute Fleiss' Kappa~\cite{fleiss1971measuring}, obtaining a score of 0.84, indicating agreement.
Second, for the clarification responses, we measure the pairwise match rate of the STL formulas generated from different participants' clarifications of the same requirement. The average pairwise match rate is 87.2\%, confirming that participants provide consistent clarifications.

\begin{table}[]
\small
\caption{Number of Requirements with Vagueness and Ambiguity Identified in the DeepSTL and STL-DivEn Datasets. ``Both'' Denotes Requirements Containing Both Vagueness and Ambiguity.}
\resizebox{0.9\textwidth}{!}{
\begin{tabular}{lcccccc}
\hline
\multirow{2}{*}{Dataset} & \multicolumn{3}{c}{Vagueness} & \multicolumn{2}{c}{Ambiguity} & \multirow{2}{*}{Both} \\ \cmidrule(lr){2-4} \cmidrule(lr){5-6}
 & Temporal & Numerical & Conditional Logic & Referential & Semantic &  \\ \hline
DeepSTL & 87 & 39 & 103 & 83 & 117 & 62 \\
STL-DivEn & 67 & 56 & 43 & 71 & 212 & 77 \\ \hline
\end{tabular}
}
\label{Table_1}
\end{table}


\begin{table*}[!t]
\centering
\caption{Metric-based Evaluation of NL-to-STL Transformation on DeepSTL and STL-DivEn. Models Are Grouped into Three Categories: Direct Transformation with Open-source Models, Direct Transformation with Closed-source Models, and Interactive Transformation. Acc. Stands for Accuracy, and Sem. Rob. denotes semantic robustness. Bold Values Indicate the Best Results; Red Values Show ClarifySTL's Improvements over the Previous SOTA.}
\resizebox{\textwidth}{!}{
\begin{tabular}{lcccc|cccc}
\hline
\multirow{2}{*}{Model} & \multicolumn{4}{c|}{DeepSTL} & \multicolumn{4}{c}{STL-DivEn} \\
\cline{2-9}
 & Formula Acc. (\%) & Template Acc. (\%) & BLEU & Sem. Rob. (\%) & Formula Acc. (\%) & Template Acc. (\%) & BLEU & Sem. Rob. (\%) \\ \hline
\multicolumn{9}{l}{{\color[HTML]{9B9B9B} \textit{Direct Transformation (Open-Source Models)}}} \\
DeepSTL & 20.02 & 29.16 & 0.3332 & 35.6 & 21.35 & 22.97 & 0.0312 & 34.7 \\
\rowcolor[HTML]{EFEFEF} 
KGST & 52.81 & 57.13 & 0.5272 & 70.3 & 57.24 & 58.83 & 0.2283 & 73.6 \\
Qwen 3-8B & 28.83 & 30.81 & 0.4121 & 44.8 & 41.74 & 43.28 & 0.0192 & 57.6 \\
\rowcolor[HTML]{EFEFEF} 
Qwen 3-14B & 37.28 & 39.21 & 0.4982 & 54.2 & 50.82 & 51.12 & 0.2092 & 66.4 \\
LLaMA 3-8B & 30.21 & 32.84 & 0.4394 & 46.7 & 44.92 & 45.73 & 0.0231 & 60.5 \\
\rowcolor[HTML]{EFEFEF} 
LLaMA 3-14B & 39.92 & 41.82 & 0.4762 & 56.4 & 53.82 & 54.92 & 0.2183 & 69.0 \\ \hdashline
\multicolumn{9}{l}{{\color[HTML]{9B9B9B} \textit{Direct Transformation (Closed-source Models)}}} \\
DeepSeek-V3 & 32.97 & 34.82 & 0.4182 & 49.1 & 51.93 & 53.72 & 0.1093 & 66.8 \\
\rowcolor[HTML]{EFEFEF} 
GPT-4o & 30.82 & 32.77 & 0.3921 & 47.8 & 55.24 & 57.93 & 0.1982 & 70.2 \\
GPT-4o-mini & 28.72 & 29.82 & 0.3728 & 43.5 & 52.83 & 53.82 & 0.1872 & 67.1 \\
\rowcolor[HTML]{EFEFEF} 
Gemini-2.5-Pro & 35.81 & 38.21 & 0.4682 & 53.1 & 56.82 & 58.32 & 0.1877 & 72.4 \\
Claude-4-Sonnet & 40.12 & 40.22 & 0.5135 & 58.4 & 55.43 & 57.93 & 0.1903 & 71.3 \\ \hdashline
\multicolumn{9}{l}{{\color[HTML]{9B9B9B} \textit{Interactive Transformation}}} \\
\rowcolor[HTML]{FFFFFF} 
DeepSeek-V3 & 38.74 & 39.82 & 0.4316 & 55.7 & 50.26 & 50.97 & 0.1221 & 63.2 \\
\rowcolor[HTML]{EFEFEF} 
GPT-4o & 34.86 & 37.64 & 0.4211 & 52.3 & 53.71 & 53.86 & 0.2046 & 67.5 \\
\rowcolor[HTML]{FFFFFF} 
GPT-4o-mini & 32.57 & 34.38 & 0.3859 & 48.6 & 49.76 & 51.68 & 0.1957 & 63.5 \\
\rowcolor[HTML]{EFEFEF} 
Gemini-2.5-Pro & 40.62 & 42.31 & 0.4826 & 58.1 & 53.67 & 55.41 & 0.2117 & 68.9 \\
Claude-4-Sonnet & 45.68 & 47.41 & 0.5162 & 63.2 & 54.67 & 55.89 & 0.2014 & 69.5 \\ \hline
\rowcolor[HTML]{EFEFEF} 
ClarifySTL & \textbf{67.12 \textcolor{red}{(14.31 $\uparrow$)}} & \textbf{70.03 \textcolor{red}{(12.90 $\uparrow$)}} & \textbf{0.5736 \textcolor{red}{(0.0464 $\uparrow$)}} & \textbf{82.7 \textcolor{red}{(12.4 $\uparrow$)}} & \textbf{70.74 \textcolor{red}{(13.50 $\uparrow$)}} & \textbf{72.28 \textcolor{red}{(13.45 $\uparrow$)}} & \textbf{0.2612 \textcolor{red}{(0.0329 $\uparrow$)}} & \textbf{84.9 \textcolor{red}{(11.3 $\uparrow$)}} \\ \hline
\end{tabular}
}
\label{Table_2}
\end{table*}

\begin{table}[!t]
\caption{Human Evaluation of Interactive Transformation Methods: STL Transformation Accuracy (\%) on Requirements with Different Types of Vagueness and Ambiguity, Evaluated on DeepSTL and STL-DivEn.}
\resizebox{\textwidth}{!}{
\begin{tabular}{clccc@{\hspace{0.8cm}}cc}
\hline
\multirow{2}{*}{Dataset} & \multicolumn{1}{c}{\multirow{2}{*}{Model}} & \multicolumn{3}{c}{Vagueness Type} & \multicolumn{2}{c}{Ambiguity Type} \\ \cmidrule(lr){3-5} \cmidrule(lr){6-7}
 & \multicolumn{1}{c}{} & \#Temporal Acc. (\%) & \# Numerical Acc. (\%) & \# Condition Acc. (\%) & \#Referential Acc. (\%) & \#Semantic Acc. (\%) \\ \hline
\multirow{6}{*}{DeepSTL} & DeepSeek-V3 & 73.8 & 75.7 & 74.2 & 78.4 & 71.3 \\
 & GPT-4o & 74.1 & 77.2 & 75.5 & 80.1 & 72.8 \\
 & GPT-4o-mini & 65.9 & 66.4 & 68.2 & 67.8 & 62.9 \\
 & Gemini-2.5-Pro & 73.7 & 76.4 & 77.3 & 80.7 & 73.1 \\
 & Claude-4-Sonnet & 74.8 & 76.3 & 78.4 & 82.4 & 74.3 \\
 & ClarifySTL & \textbf{91.5} & \textbf{95.2} & \textbf{93.7} & \textbf{96.1} & \textbf{92.8} \\ \hline
\multirow{6}{*}{STL-DivEn} & DeepSeek-V3 & 66.8 & 68.3 & 67.3 & 72.4 & 59.2 \\
 & GPT-4o & 67.3 & 69.7 & 66.3 & 73.1 & 60.6 \\
 & GPT-4o-mini & 61.8 & 62.1 & 61.5 & 68.2 & 58.3 \\
 & Gemini-2.5-Pro & 69.2 & 69.3 & 65.5 & 74.3 & 61.4 \\
 & Claude-4-Sonnet & 69.8 & 68.2 & 66.9 & 75.7 & 61.1 \\
 & ClarifySTL & \textbf{92.8} & \textbf{96.7} & \textbf{94.2} & \textbf{98.4} & \textbf{93.4} \\ \hline
\end{tabular}
}
\label{Table_3}
\end{table}

\subsubsection{Metric-Based Evaluation}
Table~\ref{Table_2} presents the metric-based evaluation results comparing ClarifySTL with other baselines. Bold numbers denote the best performance, and red values represent ClarifySTL’s relative improvements over SOTA.

We observe that ClarifySTL achieves the best performance across all evaluation datasets.
Compared with closed-source LLMs in the Direct Transformation category, it shows performance gains on both datasets. 
On DeepSTL, ClarifySTL surpasses the best-performing model, Claude-4-Sonnet, by 27.00\% in Formula Accuracy, 29.81\% in Template Accuracy, and 0.0601 in BLEU. 
On STL-DivEn, it outperforms Gemini-2.5-Pro by 13.92\% in Formula Accuracy and 13.96\% in Template Accuracy, and exceeds GPT-4o by 0.0630 in BLEU.
Compared to models in the Interactive Transformation category, ClarifySTL also shows improvements. 
On DeepSTL, it surpasses Claude-4-Sonnet by 21.44\% in Formula Accuracy, 22.62\% in Template Accuracy, and 0.0574 in BLEU. 
On STL-DivEn, ClarifySTL outperforms Claude-4-Sonnet by 16.07\% in Formula Accuracy and 16.39\% in Template Accuracy, and surpasses Gemini-2.5-Pro by 0.0495 in BLEU.
Compared with the SOTA model KGST, ClarifySTL also achieves improvements. 
On DeepSTL, it surpasses KGST by 14.31\% in Formula Accuracy, 12.90\% in Template Accuracy, and 0.0464 in BLEU.
On STL-DivEn, the improvements reach 13.50\%, 13.45\%, and 0.0329, respectively.
For semantic robustness, ClarifySTL also obtains the highest scores on both datasets, reaching 82.7\% on DeepSTL and 84.9\% on STL-DivEn. Compared with KGST, it improves semantic robustness by 12.4\% and 11.3\%, respectively. This indicates that the formulas generated by ClarifySTL not only match the reference formulas more accurately at the syntactic level, but also better preserve the expected STL satisfaction behavior on sampled signal traces.

These results demonstrate that:
(1) ClarifySTL uses requirement clarification to handle vagueness and ambiguity, achieving better performance than other methods. 
It outperforms direct transformation methods that transform natural language to STL without clarification, as well as interactive transformation approaches that rely solely on the intrinsic knowledge of LLMs.
(2) Interactive methods perform better than direct transformation. This indicates that optimizing the input natural language enables the models to generate more accurate STL formulas. 
It also suggests that in some cases, the lower accuracy of generated STL is partly caused by vague or ambiguous input, which makes it difficult for LLMs to perform correct transformations.

\subsubsection{Human Evaluation}

Table~\ref{Table_3} 
presents the human evaluation results of different interactive methods in handling various types of vagueness and ambiguity.
Taking the DeepSTL benchmark as an example, ClarifySTL achieves the highest accuracy among all methods across temporal, numerical, and conditional-logic vagueness, with accuracies of 91.5\%, 95.2\%, and 93.7\%, respectively. It also achieves 96.1\% accuracy on referential ambiguity and 92.8\% on semantic ambiguity. These results demonstrate that our framework is highly effective in handling both vagueness and ambiguity in requirements.

\begin{tcolorbox}[boxrule=1pt,colback=white,colframe=black!75,beforeafter skip=5pt, left = 1pt, right = 1pt, top = 0pt, bottom = 0pt]
\noindent\textbf{Answer to RQ1.}
ClarifySTL outperforms existing STL transformation methods, including the SOTA model KGST. On the DeepSTL dataset, it exceeds KGST by \textbf{14.31\%} in Formula Accuracy, \textbf{12.90\%} in Template Accuracy, \textbf{0.0464} in BLEU, and \textbf{12.4\%} in semantic robustness. On the STL-DivEn dataset, it outperforms KGST by \textbf{13.50\%}, \textbf{13.45\%}, \textbf{0.0329}, and \textbf{11.3\%}, respectively.
The results of semantic robustness and human evaluation further confirm its effectiveness in generating semantically reliable STL formulas.

\end{tcolorbox}

\subsection{RQ2: How effective is ClarifySTL at detecting vagueness and ambiguity in natural language requirements?}

To answer RQ2, we evaluate the ability of ClarifySTL’s Vagueness Detector and Ambiguity Detector to identify vague and ambiguous requirements. 
We compare ClarifySTL with closed-source LLM baselines, including GPT-4o, GPT-4o-mini, DeepSeek-V3, Gemini-2.5-Pro, and Claude-4-Sonnet, as well as non-expert users. The non-expert users are three computer science students who receive training on basic STL concepts and the NL-to-STL process, and are tasked with identifying vague or ambiguous requirements, and the prompting configuration are described in Section 4.5.

We evaluate the detectors on AmbiEval and on defective requirements identified from DeepSTL and STL-DivEn. To examine the effectiveness of triple-wise contrastive learning in the Ambiguity Detector, we further compare it with a pair-wise contrastive learning classifier and a fine-tuned LLaMA 3--8B classifier trained on Part B of AmbiEval. 

\subsubsection{Evaluation Results of Detection Tasks}

The evaluation results of the vagueness detection task are presented in Table~\ref{Table_4}. The Vagueness Detector outperforms baselines across all datasets and metrics. For example, on the AmbiEval dataset, it attains an accuracy of 91.1\%, a precision of 90.2\%, a recall of 91.8\%, and an F1-score of 91.0\%. Notably, non-expert users also outperform LLMs that rely solely on intrinsic knowledge, although their performance remains lower than that of the Vagueness Detector.

The evaluation results of the ambiguity detection task are shown in Table~\ref{Table_5}. 
The Ambiguity Detector also achieves the best performance across all datasets and metrics compared with baselines. 
For instance, on the DeepSTL dataset, it reaches an accuracy of 92.5\%, a precision of 91.3\%, a recall of 93.1\%, and an F1-score of 92.2\%, clearly outperforming various LLMs. 
Similarly, non-expert users also perform better than closed-source LLMs. 

These results demonstrate that both the Vagueness Detector and the Ambiguity Detector outperform models that rely solely on the intrinsic knowledge of LLMs and non-expert users in the detection tasks.

\begin{table}[!t]
\centering
\caption{Vagueness Detection Performance of Models on AmbiEval, DeepSTL and STL-DivEn. (Accuracy, Precision, Recall, F1-score Expressed as \%)}
\resizebox{\textwidth}{!}{%
\begin{tabular}{lcccccccccccc}
\toprule
\multirow{2}{*}{Model} 
& \multicolumn{4}{c}{AmbiEval} 
& \multicolumn{4}{c}{DeepSTL} 
& \multicolumn{4}{c}{STL-DivEn} \\ 
\cmidrule(lr){2-5} \cmidrule(lr){6-9} \cmidrule(lr){10-13}
 & Accuracy & Precision & Recall & F1-score 
 & Accuracy & Precision & Recall & F1-score 
 & Accuracy & Precision & Recall & F1-score \\
 \midrule
GPT-4o & 63.4 & 66.8 & 63.1 & 64.8 & 64.9 & 65.7 & 62.9 & 64.5 & 58.9 & 63.7 & 60.4 & 61.9 \\
GPT-4o-mini & 50.8 & 57.9 & 54.7 & 56.0 & 52.7 & 60.3 & 55.9 & 57.8 & 51.6 & 52.8 & 56.2 & 54.4 \\
DeepSeek-V3 & 60.1 & 65.0 & 67.2 & 66.1 & 60.9 & 68.2 & 67.9 & 68.0 & 57.8 & 61.9 & 61.6 & 61.7 \\
Gemini-2.5-Pro & 67.6 & 65.2 & 62.4 & 63.8 & 70.9 & 65.4 & 62.8 & 64.1 & 66.9 & 60.8 & 59.9 & 60.3 \\
Claude-4-Sonnet & 67.1 & 68.7 & 69.8 & 69.2 & 72.6 & 73.4 & 70.2 & 71.8 & 65.8 & 66.8 & 70.4 & 68.6 \\
Non-expert Users & 79.3 & 79.0 & 80.1 & 79.5 & 82.4 & 81.8 & 82.6 & 82.2 & 80.1 & 80.0 & 80.9 & 80.4 \\
Vagueness Detector & \textbf{91.1} & \textbf{90.2} & \textbf{91.8} & \textbf{91.0} & \textbf{92.2} & \textbf{91.4} & \textbf{91.9} & \textbf{91.6} & \textbf{89.1} & \textbf{90.0} & \textbf{90.9} & \textbf{90.4} \\
\bottomrule
\end{tabular}%
}
\label{Table_4}
\end{table}

\begin{table}[!t]
\centering
\caption{Ambiguity Detection Performance of Models on AmbiEval, DeepSTL and STL-DivEn. (Accuracy, Precision, Recall, F1-score Expressed as \%)}
\resizebox{\textwidth}{!}{%
\begin{tabular}{lcccccccccccc}
\toprule
\multirow{2}{*}{Model} 
& \multicolumn{4}{c}{AmbiEval} 
& \multicolumn{4}{c}{DeepSTL} 
& \multicolumn{4}{c}{STL-DivEn} \\ 
\cmidrule(lr){2-5} \cmidrule(lr){6-9} \cmidrule(lr){10-13}
 & Accuracy & Precision & Recall & F1-score 
 & Accuracy & Precision & Recall & F1-score 
 & Accuracy & Precision & Recall & F1-score \\
\midrule
GPT-4o & 54.9 & 52.0 & 55.7 & 54.1 & 52.8 & 51.9 & 56.8 & 54.4 & 52.7 & 54.8 & 55.9 & 55.3 \\
GPT-4o-mini & 42.6 & 47.7 & 48.7 & 48.2 & 44.7 & 45.8 & 45.7 & 45.7 & 45.8 & 42.7 & 49.7 & 45.9 \\
DeepSeek-V3 & 53.2 & 56.8 & 53.1 & 54.9 & 50.9 & 54.0 & 52.9 & 53.4 & 59.1 & 52.8 & 52.7 & 52.8 \\
Gemini-2.5-Pro & 51.2 & 52.9 & 52.3 & 52.6 & 57.1 & 55.8 & 54.9 & 55.3 & 54.0 & 58.0 & 55.1 & 56.5 \\
Claude-4-Sonnet & 54.4 & 53.1 & 54.1 & 53.6 & 53.9 & 55.0 & 55.2 & 55.1 & 57.1 & 55.8 & 57.6 & 56.7 \\
Non-expert Users & 72.0 & 71.8 & 72.4 & 72.1 & 72.9 & 72.8 & 73.4 & 73.1 & 71.9 & 72.9 & 73.3 & 73.1 \\
Ambiguity Detector & \textbf{93.1} & \textbf{92.2} & \textbf{90.8} & \textbf{91.5} & \textbf{92.5} & \textbf{91.3} & \textbf{93.1} & \textbf{92.2} & \textbf{91.2} & \textbf{91.1} & \textbf{91.5} & \textbf{91.3} \\ 
\bottomrule
\end{tabular}%
}
\label{Table_5}
\end{table}

\subsubsection{Overlapping Evaluation}

\begin{figure}[!t]
    \centering
    \includegraphics[width=0.9\linewidth]{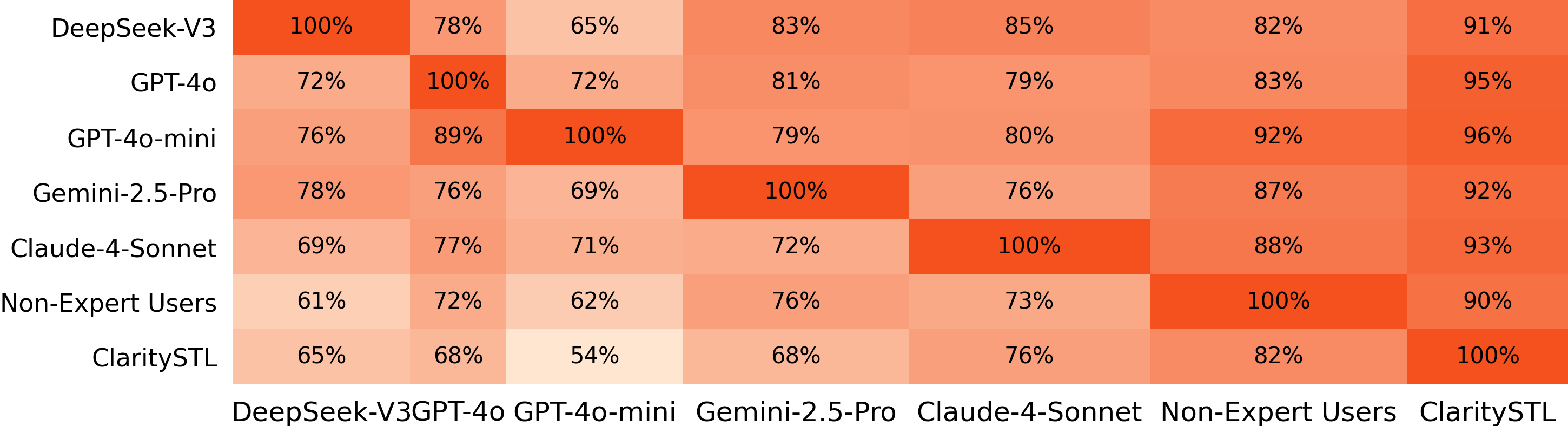}
    \caption{Overlap of Vague Requirements Detected by Different Models.}
    \label{Figure_8}
\end{figure}

\begin{figure}[!t]
    \centering
    \includegraphics[width=0.9\linewidth]{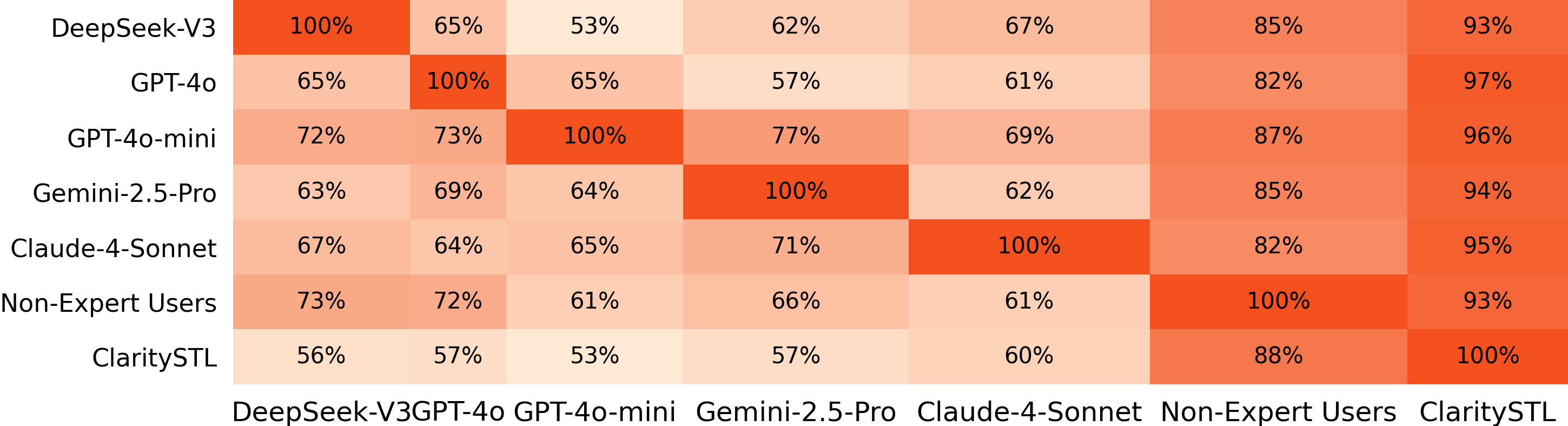}
    \caption{Overlap of Ambiguous Requirements Detected by Different Models.}
    \label{Figure_9}
\end{figure}

As illustrated in Figure~\ref{Figure_8}, each row represents the degree of overlap between the vague requirements correctly identified by one model and those identified by other models. Similarly, Figure~\ref{Figure_9} shows the overlap between the ambiguous requirements correctly recognized by different models. 
The color intensity of each rectangle deepens as the overlap increases, providing an intuitive visual cue for result interpretation. For example, 72\% of the vague requirements correctly identified by GPT-4o overlap with those recognized by GPT-4o-mini (Row 2, Column 3 in Figure~\ref{Figure_8}). 
As shown in the figures, the issues detected by other models are also recognized by our detectors, whereas the issues identified by our model are often missed by the others. This further confirms that our approach exhibits superior detection capability compared with the baselines.

\subsubsection{Contrastive Learning in Ambiguity Detection}

\begin{figure}[!t]
    \centering
    \includegraphics[width=0.5\linewidth]{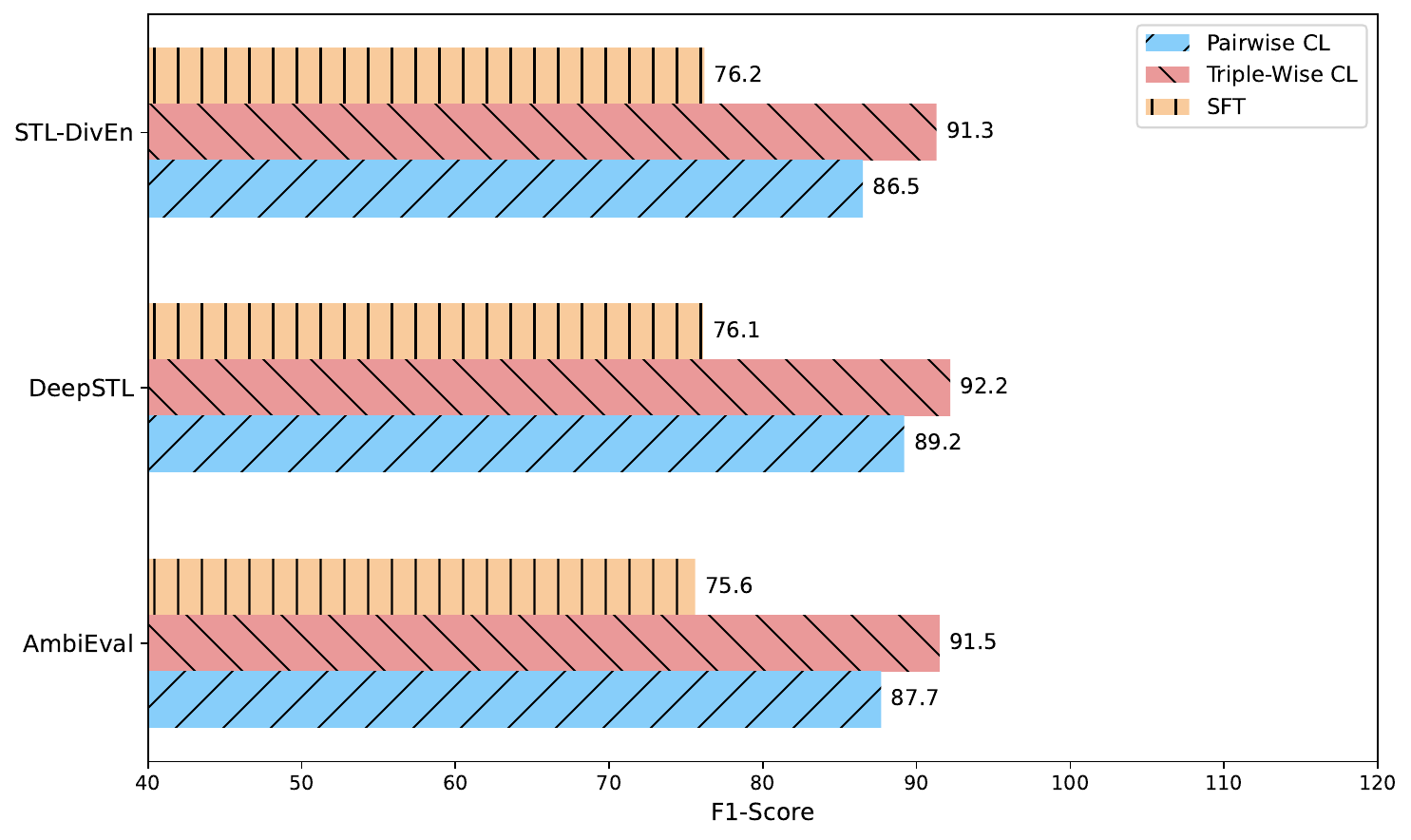}
    \caption{Comparison of Different Ambiguity Identification Methods.}
    \label{Figure_10}
\end{figure}

To validate the effectiveness of the contrastive learning module in the Ambiguity Detector, we compare three binary classification methods: triple-wise contrastive learning, pair-wise contrastive learning, and Supervised Fine-Tuning (SFT). As shown in Figure~\ref{Figure_10}, triple-wise contrastive learning consistently outperforms the other two approaches. 
This is because, during training, triple-wise contrastive learning simultaneously models the relationships among anchor, positive, and negative samples. This enables the Ambiguity Detector to establish clearer semantic boundaries for ambiguous instances.
In contrast, pair-wise contrastive learning only considers relationships between sample pairs, while SFT does not explicitly model the semantic associations among samples. 
Overall, these results demonstrate the advantage of triple-wise contrastive learning in capturing subtle semantic differences and achieving effective ambiguity detection.

\begin{tcolorbox}[boxrule=1pt,colback=white,colframe=black!75,beforeafter skip=5pt, left = 1pt, right = 1pt, top = 0pt, bottom = 0pt]
\noindent\textbf{Answer to RQ2.} ClarifySTL effectively detects both vagueness and ambiguity in requirements. Its Vagueness Detector achieves \textbf{92.2\%} accuracy on the DeepSTL dataset, \textbf{91.1\%} on AmbiEval, and \textbf{89.1\%} on STL-DivEn, with accuracy \textbf{9.8\%}, \textbf{11.8\%}, and \textbf{9.0\%} higher than the best baseline, respectively. The Ambiguity Detector, leveraging triple-wise contrastive learning, reaches \textbf{93.1\%} accuracy on AmbiEval, \textbf{92.5\%} on DeepSTL, and \textbf{91.2\%} on STL-DivEn, outperforming the best baseline by \textbf{21.1\%}, \textbf{19.6\%}, and \textbf{19.3\%} in accuracy respectively.
\end{tcolorbox}

\subsection{RQ3: Are the queries generated by ClarifySTL effective and clear for users to provide clarification?}

In this research question, we investigate the effectiveness and clarity of the clarification queries generated by ClarifySTL through a combination of metric-based and human evaluation on the AmbiEval dataset. Baseline models include GPT-4o, GPT-4o-mini, DeepSeek-V3, Gemini-2.5-Pro, Claude-4-Sonnet and non-expert users.


For human evaluation, three participants with at least two years of experience in STL specification writing are recruited, all of whom are graduate students in the CPS domain.
Each participant receives a detailed evaluation guideline that defines the criteria for correctness (whether the query accurately identifies the vague or ambiguous part of the requirement) and clarity (whether the query is easy to understand and actionable for users).
Each participant receives a set of vague or ambiguous requirements along with the corresponding generated queries.
Each query is rated for correctness and clarity, with a score of 1 indicating satisfactory and 0 indicating unsatisfactory. The final metric is the percentage of queries receiving a score of 1.
To measure inter-annotator agreement, we compute Fleiss’ Kappa across the three annotators. The Fleiss’ Kappa values are 0.89 for correctness and 0.85 for clarity, indicating agreement among the annotators.

For vagueness query generation, "default" refers to the few-shot prompting method, while "ours" denotes the CoT prompts we designed.
For ambiguity query generation, "STL-based Analysis" indicates that LLMs analyze different STL formulas to generate queries, whereas "NL-based Analysis" follows the method proposed in Section 3.5, which analyzes natural language obtained through STL back-translation to generate queries.

Furthermore, we investigate the impact of the number of candidate STL formulas on the quality of queries in Ambiguity Inquirer. 
Specifically, the queries generated using 1 to 10 examples are evaluated by calculating their ROUGE and BERTScore with the ground truth, as well as their correctness and clarity (in the case of a single candidate STL formula, the LLM analyzes the differences between its back-translated natural language and the original requirement).

\begin{table}[!t]
\caption{Vagueness Query Generation Abilities of Models on the AmbiEval Dataset.}
\resizebox{0.75\textwidth}{!}{
\begin{tabular}{llcccc}
\hline
\multicolumn{2}{c}{} &  &  & \multicolumn{2}{c}{HumanEval} \\ \cline{5-6} 
\multicolumn{2}{c}{\multirow{-2}{*}{Method}} & \multirow{-2}{*}{ROUGE} & \multirow{-2}{*}{BERTScore} & Correctness (\%) & Clarity (\%) \\ \hline
\rowcolor[HTML]{EFEFEF} 
\cellcolor[HTML]{EFEFEF} & default & 0.386 & 0.679 & 63.8 & 71.6 \\
\rowcolor[HTML]{EFEFEF} 
\multirow{-2}{*}{\cellcolor[HTML]{EFEFEF}GPT-4o} & Ours & \textbf{0.591} & \textbf{0.875} & \textbf{92.6} & 89.3 \\
 & default & 0.294 & 0.537 & 57.6 & 51.4 \\
\multirow{-2}{*}{GPT-4o-mini} & Ours & 0.486 & 0.731 & 86.8 & 82.4 \\
\rowcolor[HTML]{EFEFEF} 
\cellcolor[HTML]{EFEFEF} & default & 0.395 & 0.608 & 62.0 & 62.3 \\
\rowcolor[HTML]{EFEFEF} 
\multirow{-2}{*}{\cellcolor[HTML]{EFEFEF}DeepSeek-V3} & Ours & 0.542 & 0.861 & 88.9 & 90.4 \\
 & default & 0.360 & 0.631 & 69.0 & 68.8 \\
\multirow{-2}{*}{Gemini-2.5-Pro} & Ours & 0.583 & 0.869 & 90.6 & \textbf{92.1} \\
\rowcolor[HTML]{EFEFEF} 
\cellcolor[HTML]{EFEFEF} & default & 0.381 & 0.635 & 64.2 & 70.7 \\
\rowcolor[HTML]{EFEFEF} 
\multirow{-2}{*}{\cellcolor[HTML]{EFEFEF}Claude-4-Sonnet} & Ours & 0.574 & 0.831 & 89.1 & 90.5 \\ \hline
\multicolumn{2}{c}{Non-expert Users} & 0.545 & 0.713 & 71.6 & 90.1 \\ \hline
\end{tabular}
}
\label{Table_6}
\end{table}

\begin{table}[!t]
\caption{Ambiguity Query Generation Abilities of Models on the AmbiEval Dataset.}
\resizebox{0.85\textwidth}{!}{
\begin{tabular}{llcccc}
\hline
\multicolumn{2}{c}{} &  &  & \multicolumn{2}{c}{HumanEval} \\ \cline{5-6} 
\multicolumn{2}{c}{\multirow{-2}{*}{Method}} & \multirow{-2}{*}{ROUGE} & \multirow{-2}{*}{BERTScore} & Correctness (\%) & Clarity (\%) \\ \hline
\rowcolor[HTML]{EFEFEF} 
\cellcolor[HTML]{EFEFEF} & default & 0.281 & 0.428 & 45.6 & 70.6 \\
\rowcolor[HTML]{EFEFEF} 
\cellcolor[HTML]{EFEFEF} & STL-based Analysis & 0.401 & 0.648 & 67.1 & 85.6 \\
\rowcolor[HTML]{EFEFEF} 
\multirow{-3}{*}{\cellcolor[HTML]{EFEFEF}GPT-4o} & NL-based Analysis (Ours) & \textbf{0.637} & \textbf{0.894} & \textbf{94.6} & 90.6 \\
 & default & 0.194 & 0.388 & 41.7 & 52.0 \\
 & STL-based Analysis & 0.322 & 0.590 & 60.5 & 77.9 \\
\multirow{-3}{*}{GPT-4o-mini} & NL-based Analysis (Ours) & 0.486 & 0.785 & 90.7 & 87.6 \\
\rowcolor[HTML]{EFEFEF} 
\cellcolor[HTML]{EFEFEF} & default & 0.296 & 0.442 & 46.0 & 61.2 \\
\rowcolor[HTML]{EFEFEF} 
\cellcolor[HTML]{EFEFEF} & STL-based Analysis & 0.427 & 0.661 & 67.6 & 82.6 \\
\rowcolor[HTML]{EFEFEF} 
\multirow{-3}{*}{\cellcolor[HTML]{EFEFEF}DeepSeek-V3} & NL-based Analysis (Ours) & 0.616 & \textbf{0.894} & \textbf{94.6} & \textbf{91.4} \\
 & default & 0.259 & 0.414 & 43.6 & 67.6 \\
 & STL-based Analysis & 0.376 & 0.629 & 66.7 & 79.8 \\
\multirow{-3}{*}{Gemini-2.5-Pro} & NL-based Analysis (Ours) & 0.629 & 0.887 & 94.1 & 90.1 \\
\rowcolor[HTML]{EFEFEF} 
\cellcolor[HTML]{EFEFEF} & default & 0.287 & 0.456 & 47.5 & 71.9 \\
\rowcolor[HTML]{EFEFEF} 
\cellcolor[HTML]{EFEFEF} & STL-based Analysis & 0.419 & 0.670 & 68.6 & 79.1 \\
\rowcolor[HTML]{EFEFEF} 
\multirow{-3}{*}{\cellcolor[HTML]{EFEFEF}Claude-4-Sonnet} & NL-based Analysis (Ours) & 0.634 & 0.884 & 93.6 & 88.6 \\ \hline
\multicolumn{2}{c}{Non-expert Users} & 0.496 & 0.731 & 76.6 & 86.2 \\ \hline
\end{tabular}
}
\label{Table_7}
\end{table}

\subsubsection{Evaluation of Query Generation}
Table~\ref{Table_6} presents the evaluation results of Vagueness Inquirer in generating queries for vagueness issues. 
Overall, Vagueness Inquirer demonstrates effectiveness and clarity in generating queries. 
For metric-based evaluation, the GPT-4o-based Vagueness Inquirer achieves the best performance, with ROUGE and BERTScore values of 0.591 and 0.875 respectively. 
In human evaluation, it attains the highest correctness at 92.6\%, while the Gemini-2.5-Pro-based Vagueness Inquirer achieves the highest clarity at 92.1\%.
Meanwhile, the queries generated by non-expert users are of relatively high quality, outperforming all default baselines but still underperforming our method.

Table~\ref{Table_7} presents the evaluation results of the Ambiguity Inquirer in generating queries for ambiguity issues. For metric-based evaluation, the GPT-4o-based Ambiguity Inquirer performs best, achieving a ROUGE score of 0.637 and a BERTScore of 0.894, which are comparable to the scores of DeepSeek-V3. In human evaluation, both GPT-4o and DeepSeek-V3 attain the highest correctness at 94.6\%, while DeepSeek-V3 achieves the highest clarity at 91.4\%.
In comparison, the STL-based analysis shows a noticeable drop in both metric-based evaluation and human evaluation, performing slightly better than the default baselines but lower than non-expert users and our method. 
This indicates that leveraging LLMs for natural language analysis is more effective than directly analyzing STL formulas.


\begin{figure}[!t]
    \centering
    \includegraphics[width=0.55\linewidth]{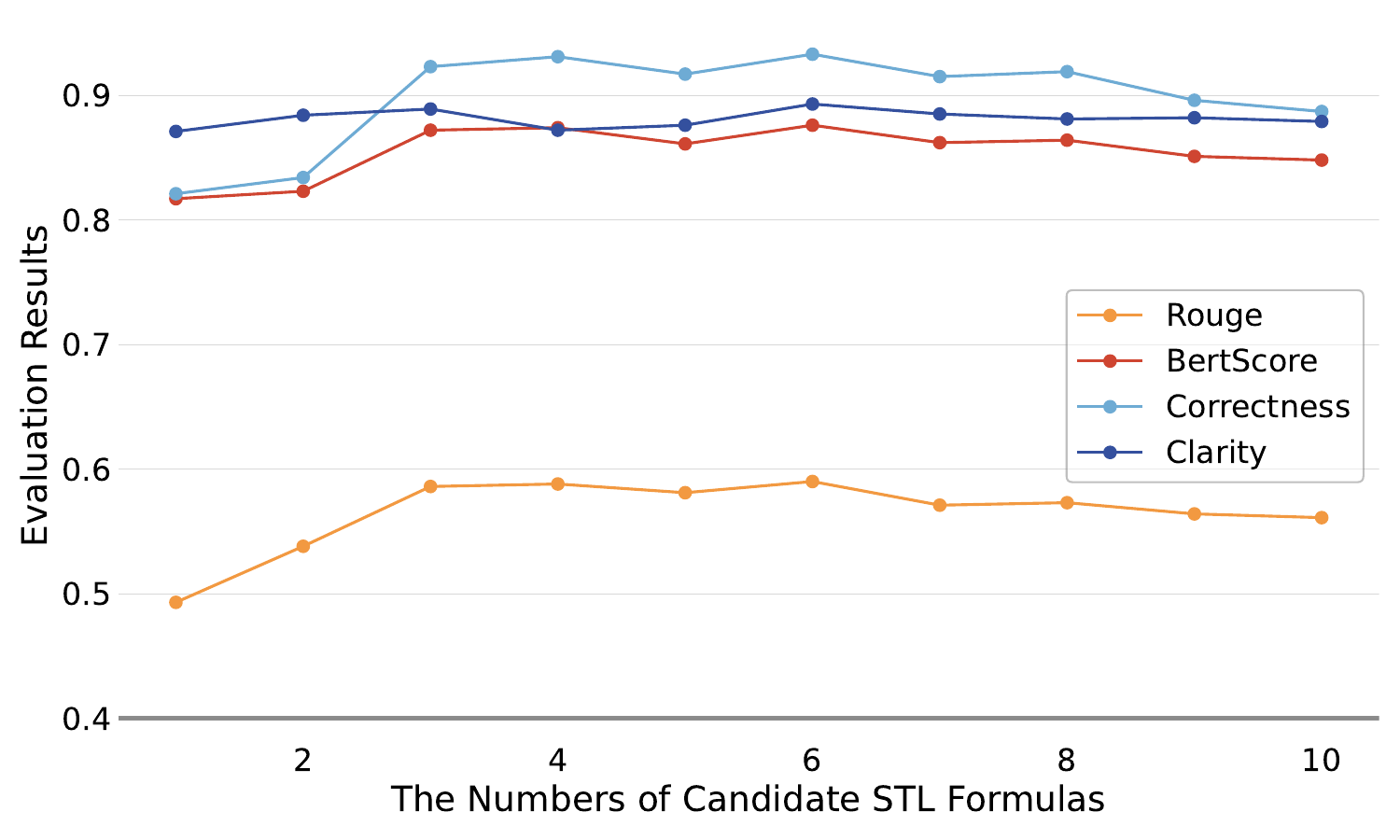}
    \caption{Impact of the Number of Candidate STL Formulas.}
    \label{Figure_11}
\end{figure}

\subsubsection{Impact of the Number of Candidate STL Formulas}
As illustrated in Figure~\ref{Figure_11}, the number of candidate STL formulas used in prompts impacts model performance.
When the number of candidate STL formulas is fewer than three, increasing their quantity enhances the model’s ability to generate high-quality queries. 
However, when the number exceeds three, the performance improvement becomes marginal, and excessive candidates even lead to a decline. This occurs because adding a few candidates provides helpful contextual cues, but too many introduce noise and redundancy, making it harder for the model to focus and thus reducing query generation quality.

\begin{tcolorbox}[boxrule=1pt,colback=white,colframe=black!75,beforeafter skip=5pt, left = 1pt, right = 1pt, top = 0pt, bottom = 0pt]
\noindent\textbf{Answer to RQ3.} 
ClarifySTL proves effective and precise in generating clarification queries. The Vagueness Inquirer achieves a ROUGE score of \textbf{0.591} and a BERTScore of \textbf{0.875}, while the Ambiguity Inquirer scores \textbf{0.637} and \textbf{0.894}, respectively. Human evaluation confirms the quality, with the Vagueness Inquirer reaching \textbf{92.6\%} correctness and \textbf{92.1\%} clarity, and the Ambiguity Inquirer achieving \textbf{94.6\%} correctness and \textbf{91.4\%} clarity, outperforming the baselines.
\end{tcolorbox}

\section{Discussion} \label{sec:discussion}

\subsection{Ablation Study}
\begin{table}[H]
\caption{Ablation Experiments Conducted on the DeepSTL Dataset. (\faCheck~indicates including the component, \faTimes~indicates removing the component. Acc. stands for accuracy, Avg. stands for average, and Sem. Rob. stands for semantic robustness.)}
\resizebox{\textwidth}{!}{
\begin{tabular}{ccccccccc}
\hline
\multicolumn{2}{c}{Vagueness Stage} & \multicolumn{2}{c}{Ambiguity Stage} & \multirow{2}{*}{Formula Acc. (\%)} & \multirow{2}{*}{Template Acc. (\%)} & \multirow{2}{*}{BLEU} & \multirow{2}{*}{Sem. Rob. (\%)} & \multirow{2}{*}{\begin{tabular}[c]{@{}c@{}}Avg. Iteration Rounds  \\ per Requirement\end{tabular}} \\ 
\cmidrule(lr){1-2}
\cmidrule(lr){3-4}
Detector & Inquirer & Detector & Inquirer &  &  &  &  &  \\ \hline
\multicolumn{2}{c}{\faCheck} & \multicolumn{2}{c}{\faCheck} & 67.12 & 70.03 & 0.5736 & 82.7 & 3.7 \\
\multicolumn{2}{c}{\faCheck} & \multicolumn{2}{c}{\faTimes} & 54.82 & 53.91 & 0.4386 & 69.4 & 3.2 \\
\multicolumn{2}{c}{\faTimes} & \multicolumn{2}{c}{\faCheck} & 53.76 & 54.83 & 0.4471 & 68.7 & 2.9 \\
\faCheck & \faTimes & \multicolumn{2}{c}{\faCheck} & 62.18 & 63.86 & 0.5218 & 77.9 & 6.5 \\
\multicolumn{2}{c}{\faCheck} & \faCheck & \faTimes & 61.27 & 61.93 & 0.5094 & 76.8 & 6.0 \\
\faTimes & \faCheck & \multicolumn{2}{c}{\faCheck} & 58.91 & 57.84 & 0.4892 & 72.6 & 2.3 \\
\multicolumn{2}{c}{\faCheck} & \faTimes & \faCheck & 56.84 & 57.35 & 0.4784 & 71.5 & 2.9 \\ \hline
\end{tabular}
}
\label{Ablation}
\end{table}

The results of the ablation experiments are shown in Table~\ref{Ablation}. Removing the Detector or Inquirer means using only the few-shot prompt to enable LLMs to detect errors or generate queries, respectively.
It can be observed that when the Inquirer is removed, the number of iterations per requirement increases, indicating higher labor costs. 
Meanwhile, the metric-based evaluation metrics decrease, which suggests that the lack of the Inquirer partially causes annotators to provide inaccurate clarification responses. 
When the Detector is removed, the automated metrics decline and the average number of iteration decreases. 
This indicates that fewer errors are detected, resulting in the early termination of the clarification process. 
Moreover, the removal of either stage leads to a drop in metric-based evaluation, demonstrating the significance of proposing both stages.


\subsection{Rationale for Selecting the Base Model}
The results in Table~\ref{Table_8} demonstrate that the performance of the detection agents is influenced by the choice of the underlying base model. 
Models with larger parameter sizes consistently outperform their smaller counterparts in both vagueness and ambiguity detection. 
Specifically, LLaMA 3-14B achieves 94.0\% accuracy and 93.2\% F1-score for vagueness detection, as well as 94.9\% accuracy for ambiguity detection.
Qwen 3-14B also performs competitively, attaining an F1-score of 92.9\% in ambiguity detection.
In contrast, smaller models such as LLaMA 3-8B and Qwen 3-8B show drops in all metrics. 
However, considering the trade-off between performance, computational efficiency, and deployment scalability, we adopt LLaMA 3-8B as the primary base model for experiments.

As shown in Table~\ref{Table_6} and Table~\ref{Table_7}, GPT-4o achieves outstanding performance across most metrics in both tasks. Therefore, we select GPT-4o as the base model for the Inquirer, considering its strong capability and reasonable computational cost.
\begin{table}[!t]
\caption{Impact of Different Base Models on the AmbiEval Dataset for Vagueness and Ambiguity Detection.}
\resizebox{\textwidth}{!}{
\begin{tabular}{lcccccccc}
\hline
\multirow{2}{*}{Base Model} & \multicolumn{4}{c}{Vagueness Detector} & \multicolumn{4}{c}{Ambiguity Detector} \\ \cmidrule(lr){2-5} \cmidrule(lr){6-9}
 & Accuracy (\%) & Precision (\%) & Recall (\%) & F1-score (\%) & Accuracy (\%) & Precision (\%) & Recall (\%) & F1-score (\%) \\ \hline
LLaMA 3-8B & 91.1 & 90.2 & 91.8 & 91.0 & 93.1 & 92.2 & 90.8 & 91.5 \\
LLaMA 3-14B & \textbf{94.0} & 92.6 & \textbf{93.8} & \textbf{93.2} & \textbf{94.9} & \textbf{93.1} & 92.8 & 92.9 \\
Qwen 3-8B  & 90.0 & 88.9 & 87.8 & 88.3 & 90.9 & 89.8 & 89.2 & 89.5\\
Qwen 3-14B & 92.9 & \textbf{92.8} & 92.6 & 92.7 & 93.9 & 92.7 & \textbf{93.2} & \textbf{92.9} \\ \hline
\end{tabular}
}
\label{Table_8}
\end{table}

\begin{figure}[t]
    \centering
    \begin{minipage}{0.49\linewidth}
        \centering
        \includegraphics[width=\linewidth]{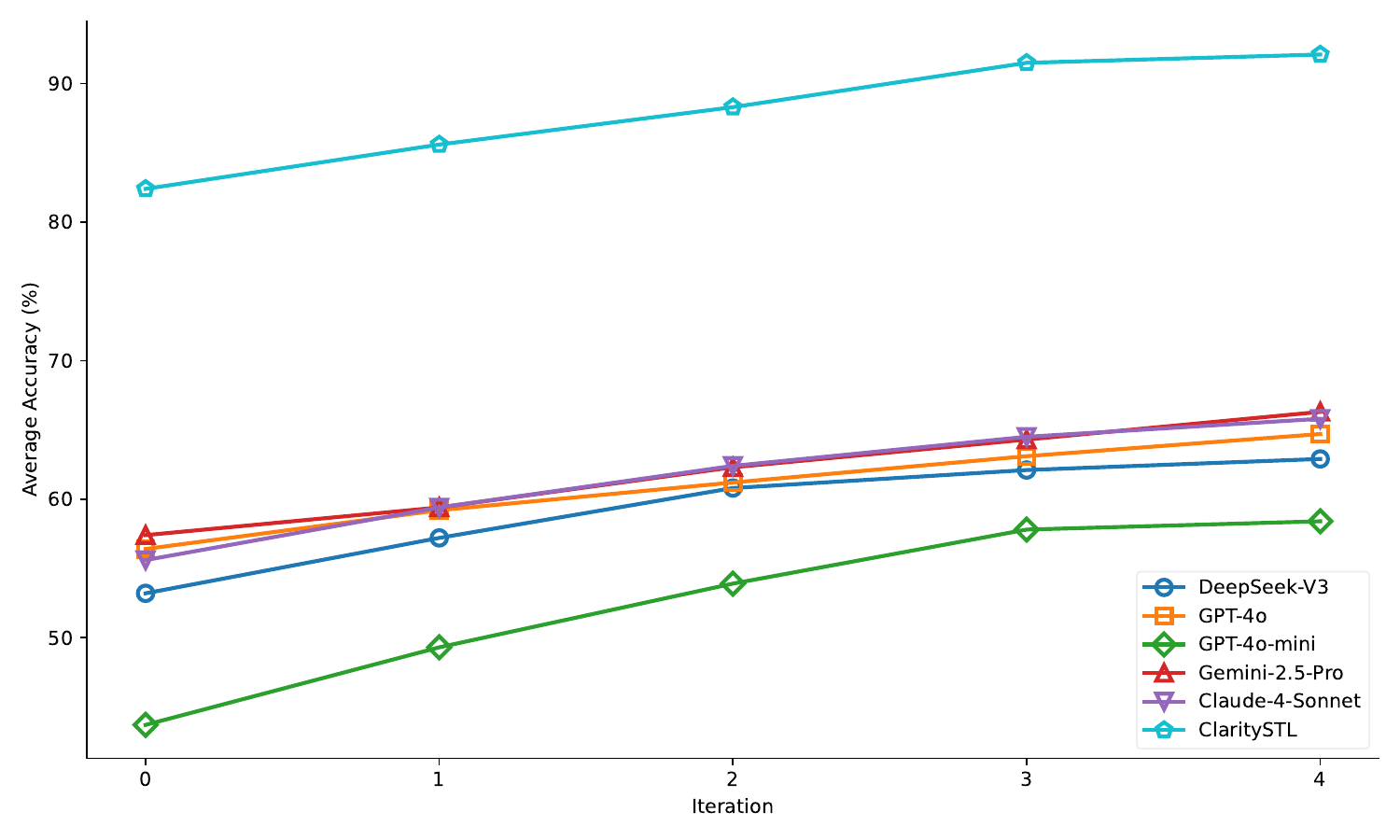}
        \caption{Impact of Iterations on Vagueness Clarification in DeepSTL.}
        \label{fig:iteration_on_deepstl}
    \end{minipage}
    \hfill
    \begin{minipage}{0.49\linewidth}
        \centering
        \includegraphics[width=\linewidth]{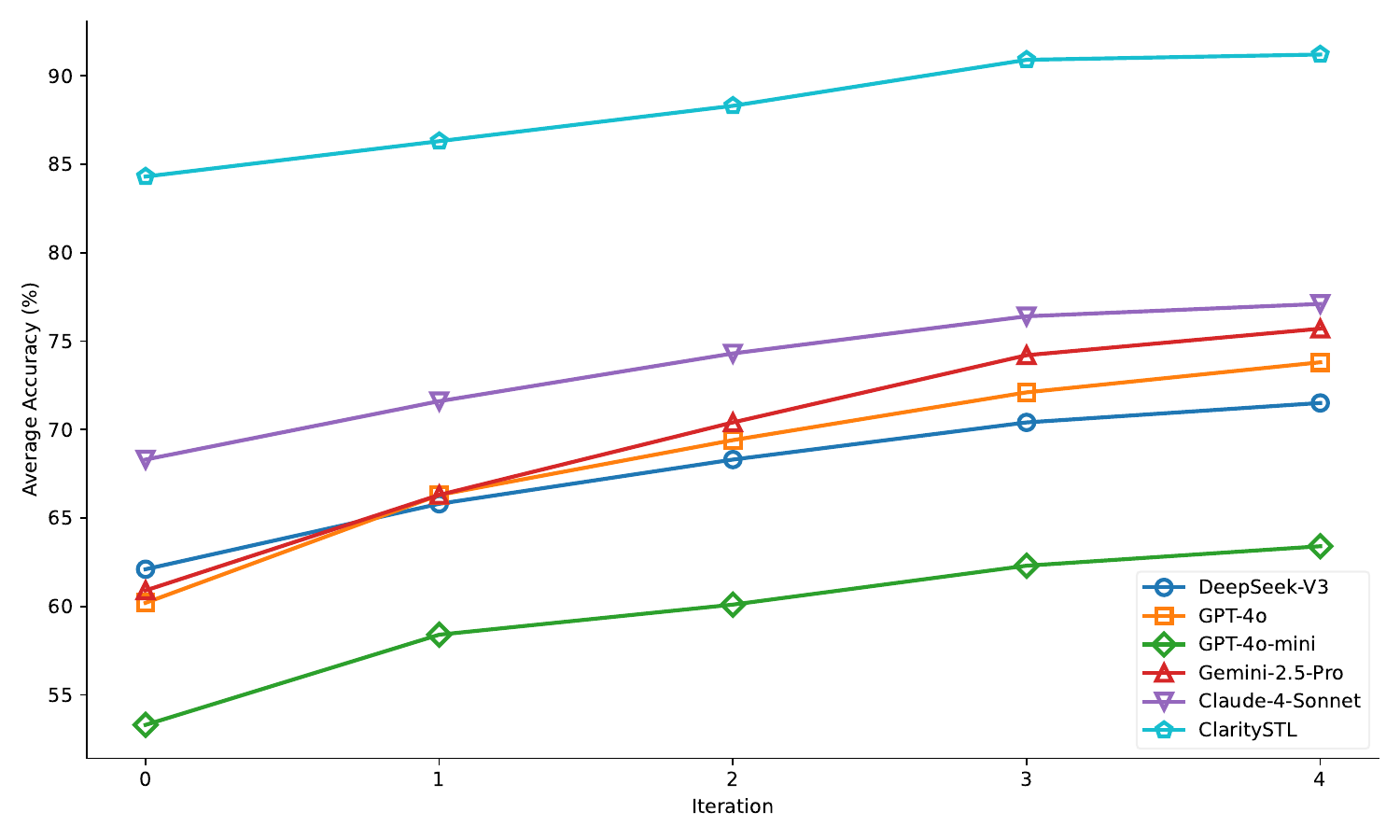}
        \caption{Impact of Iterations on Ambiguity Clarification in DeepSTL.}
        \label{fig:iteration_on_stldiven}
    \end{minipage}
\end{figure}

\subsection{Impact of Iteration Numbers}\label{sec:iteration}
To investigate the effect of the number of iterations, we measure the transformation efficiency of different models in clarifying vagueness and ambiguity on STL-DivEn as the number of iterations increases.
In our experiments, an iteration value of 0 represents the initial clarification. Additionally, the iteration number indicates the maximum number of iterations, even if the requirement still contains vagueness or ambiguity, no further iterations will be performed.
As shown in Figures~\ref{fig:iteration_on_deepstl} and~\ref{fig:iteration_on_stldiven},
the iterative structure has a positive impact on the clarification process. For example, when the iteration count is 0, ClarifySTL achieves an accuracy of 84.3\%. As the number of iterations increases, the information in the requirements is progressively supplemented or clarified, resulting in more accurate natural language requirements and, consequently, better STL formula generation. This further demonstrates that providing clearer natural language input effectively improves the accuracy of the natural language to STL transformation.

\subsection{Reliability of Ambiguity Inquirer}

The Ambiguity Inquirer involves multiple steps, including candidate STL sampling, templated back-translation, GPT-based difference analysis, and query generation. Since these steps may introduce uncertainty, we further evaluate the reliability of this pipeline from five aspects: back-translation fidelity, the influence of erroneous candidates, spurious ambiguity on unambiguous requirements, stability under repeated sampling and different temperatures, and consistency preservation after clarification. For the human evaluations in this subsection, we hire two annotators with STL experience to conduct the following assessments.

\begin{table}[t]
\centering
\caption{Back-Translation Fidelity of Candidate STL Formulas. Acc. stands for accuracy.}
\resizebox{\linewidth}{!}{
\begin{tabular}{lccccc}
\hline
Method          & Atomic Propositions Acc. (\%) & Temporal Operator Acc. (\%) & Numerical Information Acc. (\%) & Boolean Structure Acc. (\%) & Semantic Consistency (\%) \\ \hline
GPT-4o          & 95.6 & 94.8 & 94.1 & 93.9 & 95.2 \\
GPT-4o-mini     & 91.2 & 90.4 & 89.7 & 89.4 & 90.8 \\
DeepSeek-V3     & 95.1 & 94.3 & 93.8 & 93.5 & 94.9 \\
Gemini-2.5-Pro  & 94.7 & 94.0 & 93.4 & 93.0 & 94.5 \\
Claude-4-Sonnet & 94.3 & 93.7 & 93.1 & 92.8 & 94.2 \\ \hline
\end{tabular}
}
\label{tab:back_translation_fidelity}
\end{table}

\subsubsection{Back-Translation Fidelity}
To examine whether templated back-translation faithfully expresses candidate STL formulas in natural language, we conduct a human evaluation on pairs of candidate STL formulas and their back-translated descriptions. The annotators assess whether each description correctly preserves the atomic propositions, temporal operators, numerical information, Boolean structure, and overall semantics of the corresponding STL formula. A description is counted as correct only when both annotators agree.

As shown in Table~\ref{tab:back_translation_fidelity}, all models achieve high fidelity across the five aspects. GPT-4o performs best, with scores between 93.9\% and 95.6\%, while the other models also maintain generally high accuracy. These results indicate that the back-translated descriptions can provide reliable input for the subsequent difference analysis.

\begin{table}[t]
\centering
\caption{Query Quality under Erroneous Candidate Settings.}
\resizebox{0.75\linewidth}{!}{
\begin{tabular}{lcccc}
\hline
Candidates Setting & ROUGE & BERTScore & Correctness (\%) & Clarity (\%) \\ \hline
Reasonable & 0.641 & 0.897 & 95.1 & 91.8 \\
Reasonable + Erroneous & 0.635 & 0.891 & 94.3 & 91.0 \\ \hline
\end{tabular}
}
\label{tab:erroneous_candidate_query_quality}
\end{table}

\subsubsection{Impact of Erroneous Candidates}
We investigate whether erroneous candidates produced during sampling affect ambiguity clarification. Annotators divide the sampled candidate STL formulas and their back-translated descriptions into reasonable interpretations and erroneous candidates, where the latter are caused by translation or generation errors. We then compare query quality when using only reasonable candidates and when mixing reasonable and erroneous candidates.

As shown in Table~\ref{tab:erroneous_candidate_query_quality}, the two settings achieve very close results. Adding erroneous candidates only slightly decreases ROUGE from 0.641 to 0.635, BERTScore from 0.897 to 0.891, correctness from 95.1\% to 94.3\%, and clarity from 91.8\% to 91.0\%. This suggests that occasional erroneous candidates do not substantially affect ambiguity clarification, because genuine ambiguity points tend to form stable divergence patterns across reasonable candidates, while erroneous candidates usually appear as isolated noise.

\begin{table}[!t]
\centering
\caption{Sanity-Check Results of the Ambiguity Inquirer on 50 Unambiguous Requirements.}
\resizebox{0.9\linewidth}{!}{
\begin{tabular}{lccc}
\hline
Model & \# No-Ambiguity Outputs & \# Clarification Queries & \# Uncertain/Invalid Outputs \\ \hline
GPT-4o & 48 & 0 & 2 \\
GPT-4o-mini & 42 & 2 & 6 \\
DeepSeek-V3 & 47 & 1 & 2 \\
Gemini-2.5-Pro & 45 & 0 & 5 \\
Claude-4-Sonnet & 46 & 0 & 4 \\ \hline
\end{tabular}
}
\label{Table_ambiguity_sanity_check}
\end{table}

\subsubsection{Spurious Ambiguity Analysis on Unambiguous Requirements}
We conduct a sanity-check experiment to examine whether the ambiguity pipeline incorrectly triggers clarification for unambiguous requirements. Specifically, we feed 50 unambiguous requirements into the Ambiguity Inquirer and execute the full pipeline.
As shown in Table~\ref{Table_ambiguity_sanity_check}, most unambiguous requirements are correctly handled by the Ambiguity Inquirer without generating clarification queries. Across the five LLM backbones, the majority of outputs are categorized as no-ambiguity outputs, while only a small number of cases lead to clarification queries. This indicates that the pipeline does not simply treat superficial differences among sampled STL candidates as evidence of ambiguity. Instead, it generally refrains from asking users for clarification when no concrete ambiguous segment can be localized in the original requirement.

\begin{figure}[t]
    \centering
    \begin{minipage}{0.49\linewidth}
        \centering
        \includegraphics[width=\linewidth]{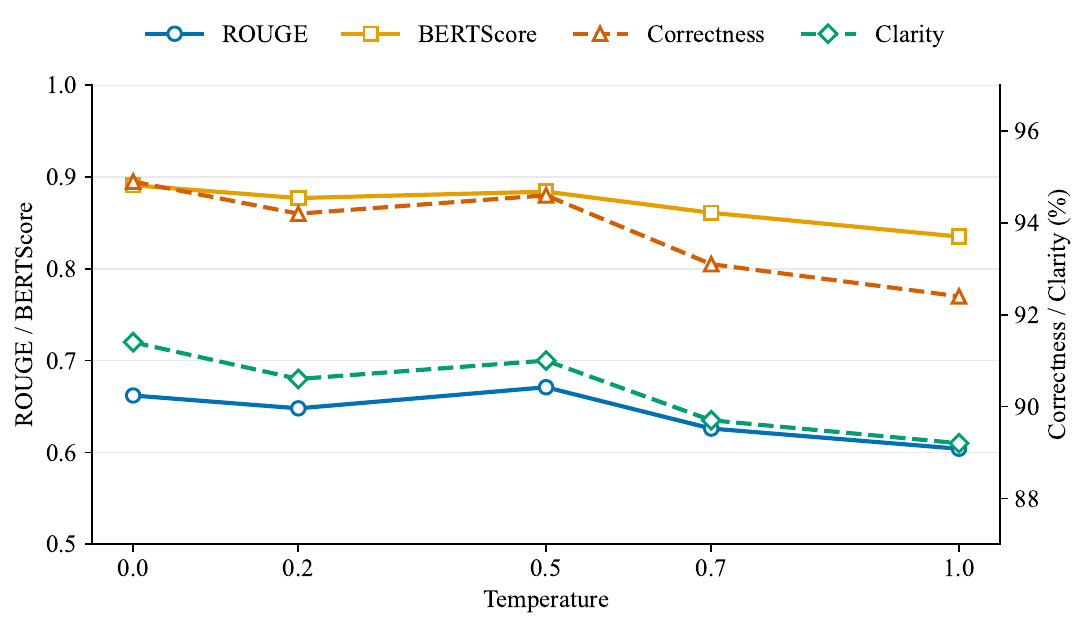}
        \caption{Temperature Sensitivity of the Ambiguity Inquirer.}
        \label{fig:temperature_sensitivity}
    \end{minipage}
    \hfill
    \begin{minipage}{0.49\linewidth}
        \centering
        \includegraphics[width=\linewidth]{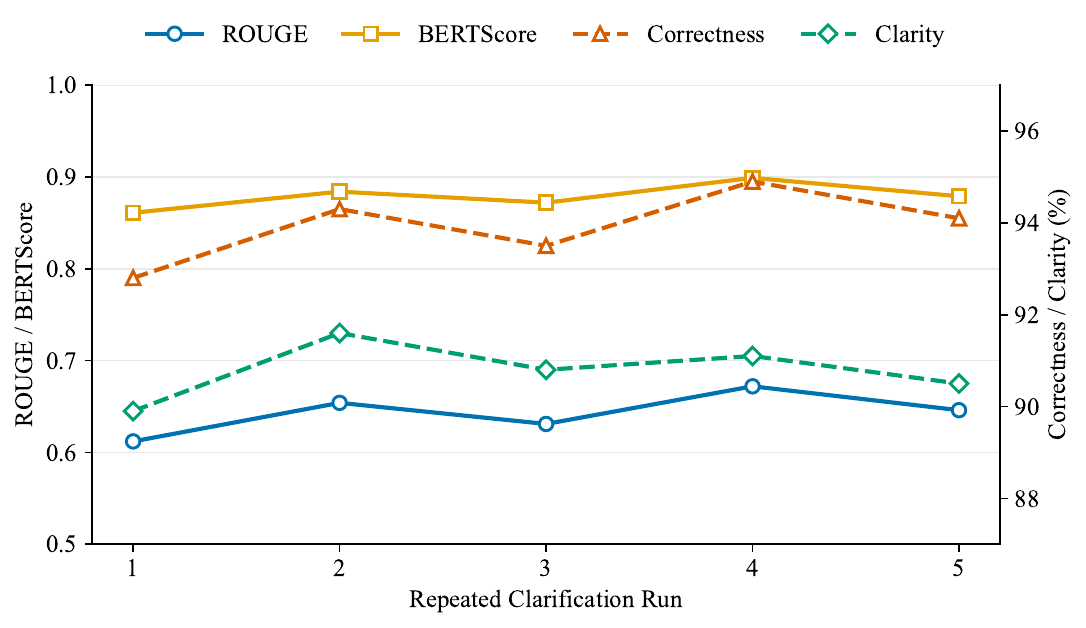}
        \caption{Performance of the Ambiguity Inquirer under Repeated Clarification Procedure.}
        \label{fig:repeated_sampling_stability}
    \end{minipage}
\end{figure}

\subsubsection{Stability and Temperature Sensitivity}
To evaluate the influence of candidate sampling and temperature settings, we repeat the ambiguity clarification pipeline for the same ambiguous requirements under multiple runs and different temperatures. 
Annotators judge whether the generated queries capture the real ambiguity in each requirement. As shown in Figures~\ref{fig:temperature_sensitivity} and~\ref{fig:repeated_sampling_stability}, the query correctness remains stable across repeated runs and stays high under different temperature settings. This indicates that the ambiguity clarification pipeline does not rely on a single accidental sampling result and is not highly sensitive to the temperature used in candidate STL generation.

\begin{table}[!t]
\centering
\caption{Alignment Preservation Between Refined and Original Requirements in Ambiguity Clarification.}
\resizebox{\linewidth}{!}{
\begin{tabular}{lcc@{\hspace{0.6cm}}cc}
\hline
\multirow{2}{*}{Model} & \multicolumn{2}{c}{With Guardrails Prompts} & \multicolumn{2}{c}{Without Guardrails Prompts} \\ \cmidrule(lr){2-3} \cmidrule(lr){4-5}
 & Requirement Consistency Rate (\%) & Avg. Clarification Rounds & Requirement Consistency Rate (\%) & Avg. Clarification Rounds \\ \hline
GPT-4o & 99.2 & 3.6 & 98.1 & 3.8 \\
GPT-4o-mini & 97.8 & 4.2 & 95.9 & 4.4 \\
DeepSeek-V3 & 99.0 & 3.7 & 98.0 & 3.9 \\
Gemini-2.5-Pro & 98.6 & 3.8 & 97.2 & 4.0 \\
Claude-4-Sonnet & 98.4 & 3.8 & 97.1 & 4.0 \\ \hline
\end{tabular}
}
\label{Table_guardrail_consistency}
\end{table}

\subsubsection{Consistency Preservation of Refined Requirements}
We further examine whether the multi-step LLM orchestration in ClarifySTL resolves ambiguity while preserving alignment with the original requirement. For each LLM backbone, we sample 50 ambiguous requirements from AmbiEval and ask annotators to evaluate whether the refined requirement correctly incorporates the user's clarification while preserving the non-ambiguous parts of the original requirement. Annotators judge whether the refined requirement preserves the original content, modifies only the user-clarified ambiguous fragment, and avoids introducing constraints not confirmed by the user. A refined requirement is counted as requirement-consistent only when all three conditions are satisfied, and the Requirement Consistency Rate is computed as the percentage of such cases. 

As shown in Table~\ref{Table_guardrail_consistency}, the consistency rate remains high across all LLM backbones, and the setting with guardrails slightly improves the rate compared with the setting without guardrails. Meanwhile, the average clarification rounds remain close under the two settings, indicating that the guardrails improve alignment preservation without increasing the user interaction burden. These results suggest that ClarifySTL’s ambiguity clarification stage is not an unconstrained rewriting process, but generally leads refinement to the user-confirmed ambiguous fragment while preserving the original natural-language specification.

\subsection{Threats to Validity}

In this subsection, we discuss the main threats to the validity of ClarifySTL as follows:

\textbf{Threats to external validity.} The primary threat to the external validity of this study lies in the quality of the experimental subjects. It remains uncertain to what extent the improvements achieved by ClarifySTL can be generalized to other STL transformation benchmarks. 
To mitigate this concern, we employ the DeepSTL and STL-DivEn datasets, consistent with prior studies. We also introduce specialized benchmark datasets for vagueness and ambiguity detection as well as query generation. This approach enhances the diversity and coverage of our evaluation.
Furthermore, the evaluation results could be influenced by the participants’ particular experience and background. To minimize this effect, we provide a standardized evaluation procedure and detailed guidelines to ensure consistency and reproducibility. 
In future work, we plan to expand the participant pool to further assess the generalizability of ClarifySTL across a broader range of user backgrounds.

Given the limited availability of openly accessible industrial STL specifications, AmbiEval is constructed from DeepSTL and STL-DivEn, two public NL-to-STL benchmarks designed to capture representative STL usage characteristics. In particular, DeepSTL samples STL formulas according to empirical operator distributions, and STL-DivEn is also constructed to reflect similar STL usage patterns. Building on these benchmarks, AmbiEval is aligned with the practical STL operator distribution and expression patterns reflected in DeepSTL and STL-DivEn.
Beyond the dataset construction, the predefined vagueness categories in ClarifySTL are derived from the information required by STL formalization, including temporal intervals, numerical thresholds, and conditional logic. Although informal or domain-specific requirements may differ in surface expressions, their underlying underspecified information often still corresponds to these STL-relevant elements. Therefore, the prompt-based Inquirer can generalize to requirements whose vagueness falls within these STL-grounded categories. Nevertheless, requirements involving highly domain-specific assumptions or defect types may require additional category design or domain adaptation, which we will address in future work.

\textbf{Threats to internal validity.} 
LLMs are highly sensitive to prompts and hyperparameters, especially the number of task examples and natural language instructions, which can significantly affect model performance. To minimize such variability, we conduct an experimental analysis on the number of examples used in ClarifySTL and select the optimal number in all experimental settings. We also maintain consistent prompts and hyperparameter settings in the baseline models.
Another potential threat is that during the process of generating queries for ambiguity, we rely on GPT-4o to generate candidate STL formulas. 
According to Table~\ref{Table_2}, this process may introduce new errors, which are in fact identified by the framework as ambiguities requiring clarification.
When additional information is received from the user, the quality of the STL formulas generated by GPT-4o improves, eventually eliminating these errors.
This also demonstrates the effectiveness of our approach: the clearer the input natural language requirement, the more accurate the generated STL formulas.

A potential concern is whether multiple vagueness issues coexisting in a single requirement may cause context loss during clarification. As described in Section~\ref{sec:approach}, the Vagueness Detector and Inquirer operate iteratively. In each round, the Detector identifies the remaining vagueness in the current requirement, the Inquirer generates a targeted query, and the refined requirement is sent back to the Detector for the next round, while the original requirement is preserved as the main context across rounds. The effectiveness of this iterative mechanism is further empirically supported in Section~\ref{sec:iteration}, which shows that the clarification quality improves as the number of iterations increases. Therefore, multiple coexisting vagueness issues can be progressively resolved through the iterative interaction rather than handled in a single round.

\textbf{Threats to construct validity.} This paper uses Formula Accuracy, Template Accuracy, and BLEU to evaluate the accuracy of the generated STL formulas. Although these metrics cannot fully replace human judgment, they provide a rigorous and objective quantitative measure, enabling rapid assessment of model performance. 
In addition, we conduct human evaluations to verify the results produced by the model across different types of natural language requirement correction tasks.
For the evaluation of query generation quality in Section 5.3, we adopt the widely used ROUGE and BERTScore metrics from text generation tasks to ensure fair comparisons. 
Future work will incorporate additional human evaluations to further validate model performance.
Moreover, ClarifySTL focuses on leveraging agents to address vagueness or ambiguity in single-sentence natural language requirements. 
In the future, we plan to extend the design components of ClarifySTL to documents containing multiple requirement statements and apply these extended components to a broader range of scenarios.
\section{Related Work} \label{sec:related_work}

In this paper, we propose a new model for transforming natural language into signal temporal logic, which is an extension of temporal logic. Thus, our work mainly relates to two research areas: (i) from NL to TL, and (ii) from NL to STL. 
In this section, we summarize related work in these two areas.

\subsection{From Natural Language to Temporal Logic} 
Some works have investigated the transformation of natural language into Temporal Logic specifications~\cite{buzhinsky2019formalization,fuggitti2023nl2ltl,xu2024learning,mendoza2024translating,shukla2025gray,englishgrammar,bombieri2023mapping}.
For instance, a catalog of temporal logic formulas that capture common specification patterns in the design of concurrent and reactive systems was proposed in~\cite{dwyer1999patterns}.
Controlled English has also been transformed into TL through the use of syntactic and grammatical dependency parsing, together with predefined mapping rules~\cite{vzilka2010temporal, santos2018formal}.
~\citet{vzilka2010temporal,santos2018formal} transform controlled English into LTL by applying syntactic and grammatical dependency parsing combined with predefined mapping rules.
Similarly, the ARSENAL framework~\cite{ghosh2016arsenal} uses NLP techniques such as n-gram analysis and dependency parsing to transform natural language requirements into formal specifications, including TL.
Although these methods are effective in specific domains, they rely on handcrafted rules and restricted language inputs, which limit their ability to handle more diverse and complex expressions. 
To overcome these limitations, the nl2spec method~\cite{CoslerHMST23} combines human feedback with LLMs’ ability to generate LTL formulas, which relies heavily on the user's professional knowledge.
In addition, NL2TL~\cite{ChenGZF23} mitigates the reliance on rule design by fine-tuning a T5 model on NL–TL pairs generated by LLMs. 
However, these TL-focused methods are not readily extended to STL, because STL introduces real-valued signals and continuous-time constraints, which pose challenges not typically addressed by traditional TL.

\subsection{From Natural Language to Signal Temporal Logic} 
As an extension of TL that incorporates real-valued dense-time signals, STL has gained widespread use in both academia and industry to meet the requirements of CPS~\cite{MadsenVSVDWDB18,mao2024nl2stl,chen2023stl}. 
Consequently, numerous efforts have been made to transform natural language into STL~\cite{li2023learning,patel2019learning,HeBNIG22,ChenGZF23,Enhancing}. 
For example, DeepSTL~\cite{HeBNIG22} trains a Transformer model using grammar-based synthetic data. 
Although this approach ensures formal consistency, it heavily relies on handcrafted rules and artificial data, which fail to capture the diversity in real-world natural language. 
As a result, its generalization capability is limited when applied to open-domain inputs. 
To facilitate broader evaluation, STL-DivEn~\cite{Enhancing} has been introduced for NL-to-STL translation. This dataset provides diverse linguistic expressions and complex signal patterns, enabling standardized assessment of model robustness and semantic fidelity. 
KGST~\cite{Enhancing} uses a generate-then-refine approach by first fine-tuning LLMs to generate initial STL formulas, then refining them with external knowledge. 
Despite their impressive performance, DeepSTL and KGST, as supervised learning approaches, rely on fixed training objectives for supervision. This results in limited mechanisms for detecting or resolving ambiguities in natural language input. Consequently, when the input contains vague or ambiguous information, these approaches struggle to accurately capture the intended semantics.
Dialogue-based approaches~\cite{mohammadinejad2024systematic} mitigate this issue by leveraging the internal knowledge of LLMs to engage in dialogue with users regarding unclear parts, refining requirements for STL generation.
However, the internal knowledge of LLMs is typically limited, making it difficult for them to identify vague or ambiguous information expressed in requirements and accurately generate queries to interact with users.
The objective of ClarifySTL is to provide a novel interactive framework: we inject vagueness classification knowledge through fine-tuning and design CoT-based prompts to clarify vagueness. Additionally, we capture the subtle differences between ambiguous and unambiguous sentences through contrastive learning, and analyze the multiple potential interpretations of the same input requirement to resolve ambiguity.
Experimental results demonstrate that the proposed detection and query generation methods enable ClarifySTL to conduct efficient interactions and requirement clarifications, generating STL specifications that accurately reflect user intent.

\section{Conclusion and Future Work} \label{sec:conclusion}
This paper introduces an interactive NL–to-STL transformation framework that guides users in clarifying vagueness and ambiguity in natural language requirements, helping users transform such requirements into STL formulas that align with their intended semantics.
ClarifySTL is built upon LLMs fine-tuned with AmbiEval and GPT-4o, enabling it to accurately identify vague or ambiguous requirements and generate explicit queries.
Metric-based evaluations on the DeepSTL, STL-DivEn, and AmbiEval benchmarks demonstrate that ClarifySTL outperforms baselines.
Human evaluations further validate the effectiveness of our framework, showing that it effectively guides users in clarification.
We also conduct dedicated experiments to evaluate the accuracy of our vagueness and ambiguity detection agents, as well as the correctness and clarity of the corresponding generated queries.
In future work, we plan to explore cross-domain adaptation and extend ClarifySTL to document-level STL transformation tasks to further enhance its applicability.

\bibliographystyle{ACM-Reference-Format}
\bibliography{sample-base}  
\end{document}